\documentclass[sigconf,balance=false]{acmart}
\usepackage{popets}

\setcopyright{popets}
\copyrightyear{YYYY}

\acmYear{YYYY}
\acmVolume{YYYY}
\acmNumber{X}
\acmDOI{XXXXXXX.XXXXXXX}
\acmISBN{}
\acmConference{Proceedings on Privacy Enhancing Technologies}
\usepackage{amsmath}
\usepackage{algorithmic}
\usepackage{graphicx}
\usepackage{textcomp}
\usepackage{bmpsize}
\usepackage{xcolor}
\usepackage{lipsum}
\usepackage{enumitem}

\usepackage{caption}
\usepackage{subcaption}
\usepackage{mdframed}
\usepackage{tcolorbox}
\usepackage{tikz}
\usepackage[table]{xcolor}
\usepackage{filecontents}
\usepackage{listings}
\usepackage{multirow} 
\usepackage{adjustbox}
\usepackage{soul}
\usepackage{makecell}

\begin{document}

\title[]{The Role of Online Forums in Developer Understanding of Privacy Law - A Reddit Case Study}


\author{Sara Haghighi}
\orcid{0009-0000-8296-0475}
\affiliation{%
  \institution{University of Maine}
  \city{Orono}
  \state{Maine}
  \country{USA}}
\email{sara.haghighi@maine.edu}

\author{Clark LaChance}
\affiliation{%
  \institution{University of Maine}
  \city{Orono}
  \state{Maine}
  \country{USA}}
\email{clark.lachance@maine.edu }

\author{Ali Pourghasemi Fatideh}
\affiliation{%
  \institution{University of Maine}
  \city{Orono}
  \state{Maine}
  \country{USA}
}
\email{ali.pourghasemi@maine.edu}

\author{Travis Breaux}
\affiliation{%
 \institution{Carnegie Mellon University}
 \city{Pittsburgh}
 \state{Pensylvania}
 \country{USA}}
\email{tdbreaux@andrew.cmu.edu}

\author{Sepideh Ghanavati}
\affiliation{%
  \institution{University of Maine}
  \city{Orono}
  \state{Maine}
  \country{USA}}
\email{sepideh.ghanavati@maine.edu}


\renewcommand{\shortauthors}{Haghighi et al.}

\begin{abstract}
Software practitioners use online forums to navigate complex and often ambiguous legal privacy requirements, yet little is known about their professional backgrounds, what challenges they face, and how they use and assess the credibility of the advice received, or how they resolve ambiguities in posts. We report the findings of a survey of 223 Reddit users from regulatory-focused subreddits, complemented by a qualitative analysis of 2,248 posts and responses. Our results show that, despite holding privacy-related certifications, most participants frequently use forums to seek legal advice. Key challenges reported or identified include implementing a data protection impact assessment, reporting a data breach, and obtaining cookie consent. Reddit users often assess credibility by reviewing respondents' post history, verifying sources cited, trusting advice from recognized experts, and following up for clarity before responding. We highlight research and educational directions to bridge gaps in support needed for regulatory compliance guidance. 
\end{abstract}


\maketitle

\section{Introduction}
\label{sec:intro}

Comprehensive privacy regulations, such as the EU General Data Protection Regulation (GDPR)~\cite{GDPR}, require organizations to adopt accountable privacy-preserving practices, with significant consequences for non-compliance. Translating these regulations into concrete technical and organizational practices, however, is often complex and ambiguous. To navigate this uncertainty, developers rely on a range of strategies, from consulting legal counsel to seeking informal guidance in online communities \cite{prybylo2024evaluating}. 

Prior research shows that, particularly in smaller companies, developers may turn to platforms such as Reddit or Stack Overflow for privacy-related advice~\cite{tahaei2020understanding, prybylo2024evaluating, li2021developers, parsons2023understanding}. 
These studies report on the types of challenges developers encounter and the advice they receive. 
However, their focus is mainly limited to privacy concepts in app development, e.g., mobile permission requests, sentiment analysis, data collection and sharing practices, privacy-by-design solutions, and overlook broader compliance discussions (e.g., how to handle data subject access requests or cross-border data transfer) that take place in regulatory-focused communities 
(e.g., \texttt{r/gdpr} or \texttt{r/europrivacy}). Furthermore, these studies analyze only posts and comments; thus, they do not capture contextual factors such as participants' demographic backgrounds, how these shape their engagement and impact their perceptions and challenges, and how they apply received advice, ensure its credibility, and resolve ambiguities in practice. Hence, it remains unclear how informal interpretations of privacy regulations develop and affect compliance. 

We address these gaps 
by conducting two complementary studies: (1) a quantitative survey of 223 Reddit users who participate in regulatory-focused subreddits; and (2) a qualitative large-scale analysis of 2,248 posts related to privacy and software development from the same subreddits. 
For the large-scale analysis, we use in-context learning with both open-source and proprietary large language models (LLMs), 
to identify posts relevant to both privacy and software development and to classify discussions that describe concrete GDPR compliance challenges. 
To ensure reliability at scale, we combine automated classification with human-in-the-loop validation. We further integrate self-reported behavior and motivations collected through our survey with the posts' analysis to compare perceived challenges with those discussed online, aiming to better understand how practitioners reason about and act on informal guidance. We also explore how factors such as professional roles influence users' motivations, perceptions of challenges, assessment of the credibility of shared advice, and how they resolve ambiguities. Lastly, we demonstrate how credibility factors best predict Reddit users' decision-making behavior (i.e., whether to follow or apply received replies in their work). 
Our research questions are: 
\begin{itemize}[leftmargin = *]
     \item \textbf{RQ1:} What motivates users to engage in privacy and legal discussions on regulatory-focused subreddits, and does this vary based on professional and organizational background?
     \item \textbf{RQ2:} What are the most common challenges users face when seeking compliance or privacy-related advice, and how do they vary by professional and organizational factors?
    \item \textbf{RQ3:} How do users assess the credibility of information shared, resolve ambiguity in posts, and how do these assessments affect their decision-making processes? Does perceived credibility vary based on professional background?
    \item \textbf{RQ4:} How do the challenges reported by Reddit users compare with those identified in the posts?
\end{itemize} 

Our survey results show that most participants 
are from North America, despite the EU-focused subreddits, and hold privacy-related certifications. 
We find that user motivations vary by privacy-related background; specifically, those with a privacy-related certification or experience are more motivated to share their knowledge, while others want to seek advice or stay informed \emph{(i.e., RQ1)}.
Although most reported having in-house legal counsel, the majority occasionally seek external privacy or legal advice. Survey users reported struggling most with data protection impact assessment (DPIA) and data breach notification, while Reddit post analysis showed consent as the most prominent challenge \emph{(i.e., RQ2 \& RQ4)}. 
We identified six additional challenges in the post analysis that were not reported in the survey, e.g., handling user consent, data processing agreements, and data transfer to a third country \emph{(i.e., RQ4)}. Post analysis showed that responders often referred to guidelines such as EDPB and ICO, while the survey revealed they mostly used the NIST frameworks. These findings underscore the importance of complementary studies in capturing challenges and practices in real-world settings. 
Users rely on multiple strategies, such as checking the poster's profile and engagement history and cross-checking responses with external sources to assess credibility. 
However, some users report acting on advice with little or no verification, highlighting potential pathways to non-compliance. This concern is further underscored by contradictory responses identified on Reddit, which require additional resources to verify their accuracy \emph{(i.e., RQ3)}. 


Based on our findings, we outline future research and educational directions as well as implications for industry and regulatory bodies. 

\section{Related Work}
\label{sec:related}

Recent studies investigate developer behavior, practices, and experiences in privacy engineering activities through large-scale surveys  \cite{prybylo2024evaluating, liang2024large,alomar2022developers, serafini2024engaging, tahaei2021deciding}, or targeted interviews 
\cite{Hadar2018, horstmann2024those, horstmann2025sorry, tahaei2021privacy, alomar2022developers, dalela2021mixed, iwaya2023privacy}.  

Tahaei et al.~\cite{tahaei2021privacy, tahaei2021deciding} and Horstmann et al.~\cite{horstmann2024those} examine privacy challenges through interviews and programming studies, identify issues like poor privacy culture and communication barriers as the main challenges, and emphasize the importance of ``privacy champions''. The presence of a privacy officer and privacy education are shown to increase privacy awareness and the adoption of privacy-enhancing tools \cite{prybylo2024evaluating}. In some cases, developers were found to not think about privacy early in development activities, or to consider it simply a security concept \cite{Hadar2018, prybylo2024evaluating}. In a privacy-related programming study with 30 developers, Horstmann et al.~\cite{horstmann2025sorry} found that they rarely consider privacy in their code, and the majority failed to have privacy-compliant solutions.
Finally, studies highlight developer privacy concerns~\cite{li2018coconut}, challenges in regulatory compliance~\cite{alomar2022developers,tahaei2022charting}, and that they may prioritize app store requirements over laws, relying on the stores to flag privacy issues~\cite{alomar2022developers,tahaei2022embedding}. 

Other works focus on forum analysis to uncover developers' security and privacy concerns. Tahaei et al. \cite{tahaei2020understanding,TahaeiLiVaniea+2022+114+131,tahaei2022privacy} explored privacy issues on Stack Overflow,
and found that most concerns involve permissions, authentication, and third-party libraries. Their results show that Google and Apple Privacy labels increased the number of privacy questions, particularly regarding consent 
\cite{tahaei2020understanding}. 
Li et al. \cite{li2021developers} and Parsons et al. \cite{parsons2023understanding} examined \texttt{r/androiddev}, \texttt{r/iOSProgramming}, and \texttt{r/webdev} and found that privacy discussions are infrequent and often motivated by regulatory pressures or app store policies. They focused on trends and sentiments, rather than compliance challenges. Kumar et al.~\cite{kumar2025privacy} conducted sentiment and temporal analyses of privacy discussions across various mental health subreddits and found that privacy discussions increased significantly over time, and sentiment varied depending on the privacy themes. 

Some studies examine trust and credibility in Reddit discussions using large-scale computational approaches, analyzing how public perceptions evolve, how credibility is assessed, and how users interact across different  groups~\cite{pessianzadeh2025generative, gefen2024evolving, amini2025news}. However, these studies focus on general attitudes toward technologies (e.g., AI systems) rather than GDPR-specific compliance challenges, and do not combine survey-based insights with large-scale post analysis.

\begin{table}[t]
\centering
\footnotesize
\caption{Comparison of Related Work and Our Study}
\footnotesize{
PS = Platform Studied,
US = User Study,
MR = Mixed Roles,
GM = GDPR Mapping,
TRI = Triangulation.
Dev = Only Developers
}
\renewcommand{\arraystretch}{1.1}

\begin{tabular}{llcccc}
\toprule
\textbf{Literature} & \textbf{PS} & \textbf{US} & \textbf{MR} & \textbf{GM} & \textbf{TRI} \\
\midrule

Li et al.~\cite{li2021developers} & r/androiddev & $\times$ & $\times$ & Partial & $\times$ \\
Parsons et al.~\cite{parsons2023understanding} & Reddit mobile+web & $\times$ & $\times$ & Partial & $\times$ \\
Santos et al.~\cite{santos2024patterns} & r/gdpr (7 posts) & $\times$ & $\times$ & $\times$ & $\times$ \\
Tahaei et al.~\cite{tahaei2020understanding,tahaei2022privacy,TahaeiLiVaniea+2022+114+131} & Stack Overflow & $\times$ & $\times$ & Partial & $\times$ \\
Horstmann et al.~\cite{horstmann2024those, horstmann2025sorry} & Interviews + Survey & $\checkmark$ & Dev & $\checkmark$ & $\times$ \\
Serafini et al.~\cite{serafini2024engaging} & Survey & $\checkmark$ & Dev & $\times$ & $\times$ \\
Kumar et al.~\cite{kumar2025privacy} & Reddit Mental Health & $\times$ & $\times$ & Partial & $\times$ \\
Prybylo et al.~\cite{prybylo2024evaluating} & Survey & $\checkmark$ & $\checkmark$ & $\times$ & $\times$ \\
Pessianzadeh et al.~\cite{pessianzadeh2025generative} & Reddit discussions & $\times$ & $\checkmark$ & $\times$ & $\times$ \\
Amini et al.~\cite{amini2025news} & Reddit discussions & $\times$ & $\times$ & $\times$ & $\times$ \\

\midrule
\textbf{This Work} & \begin{tabular}[c]{@{}c@{}}Regulatory-focused\\subreddits + Survey\end{tabular}  & $\checkmark$ & $\checkmark$ & $\checkmark$ & $\checkmark$ \\

\bottomrule
\end{tabular}

\vspace{2mm}
\label{table:comparison}
\end{table}

Prior work primarily analyzes the types of questions on developer-centric platforms, like Stack Overflow, or subreddits, e.g., \texttt{r/webdev} or \texttt{r/androiddev} \cite{tahaei2020understanding,TahaeiLiVaniea+2022+114+131,tahaei2022privacy, li2021developers, parsons2023understanding}, focusing on narrow topics (e.g., mobile app permissions, personal data) or conducting sentiment and trend analysis, rather than identifying specific GDPR compliance challenges such as record of processing activities. 
As shown in Table~\ref{table:comparison} and Figure \ref{fig:methodologyfigure}, we complement these studies by analyzing regulatory-focused subreddits (e.g., \texttt{r/gdpr}, \texttt{r/europrivacy}) through a dual study. First, we survey 223 users (including developers as well as legal professionals, privacy officers, and end-users) to uncover the types of GDPR concerns raised, their engagement on Reddit, the strategies for assessing credibility in anonymous environments and resolving ambiguities, and their decision-making processes. Second, we conduct a large-scale analysis of posts using an LLM-based approach to identify GDPR challenges in software development. This combination bridges technical and legal perspectives and offers a more comprehensive view of how diverse actors grapple with GDPR compliance in practice. 
To the best of our knowledge, our work is the first study to directly engage with Reddit users to explore GDPR-related challenges and motivations as articulated by them within their communities.

\section{Study Design and Reddit Posts Analysis}
\label{sec:study}

\begin{figure*}[t]
    \centering
    \includegraphics[width=0.75\textwidth]{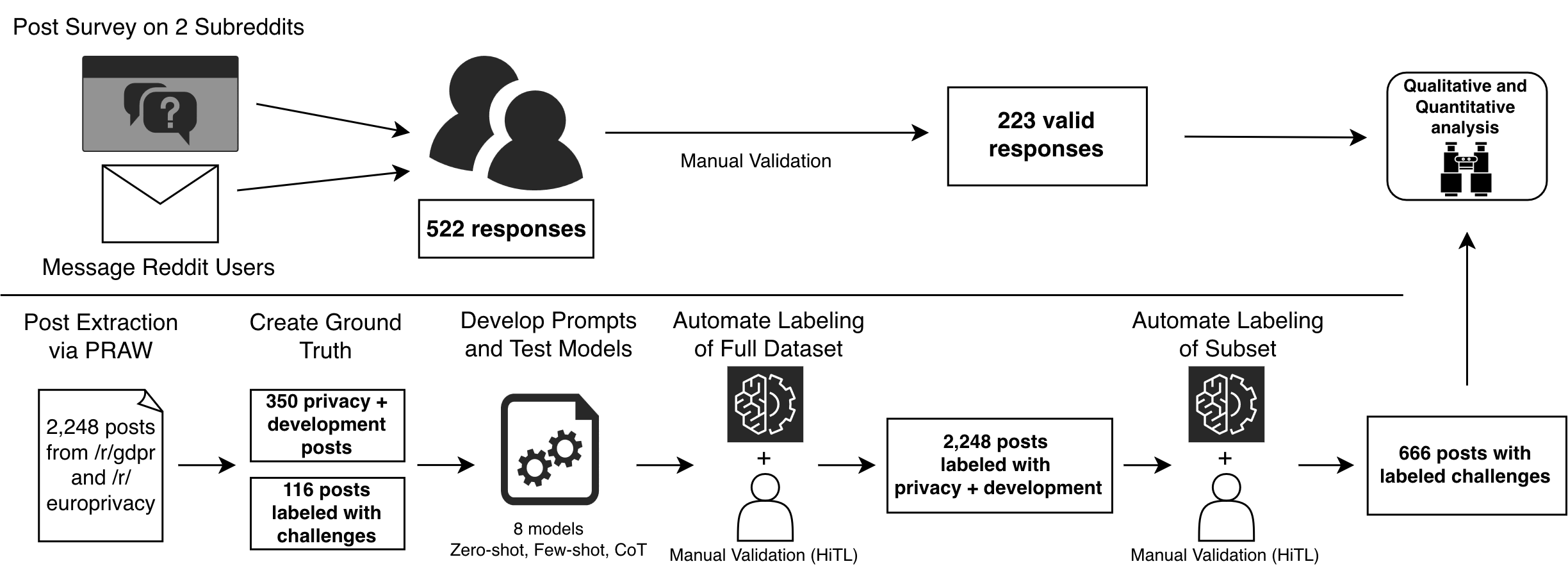}
    \caption{The Overview of Our Study}
    \label{fig:methodologyfigure}
\end{figure*}

We now present the survey design and Reddit post analysis method. Figure \ref{fig:methodologyfigure}
shows our study overview.

\subsection{Survey Design}
\label{sec:survey_design}

The survey was informed by prior research on developer privacy practices~\cite{tahaei2021privacy,horstmann2024those,prybylo2024evaluating,tahaei2019survey}, how they reason about privacy requirements on forums~\cite{santos2024patterns,li2021developers,parsons2023understanding}, and by informal discussions with our legal collaborators (i.e., law professors) and a privacy research professional at a major company. We iteratively piloted the survey with members of our university's privacy research group to improve clarity and flow. Based on feedback and pilot response analysis, we refined question wording, clarified terminology, and adjusted Likert-scale options to better align with GDPR compliance constructs identified in prior literature and the CIPL report~\cite{CIPL_GDPR_Implementation_2017}.

The survey was designed with three goals: to reveal the professional backgrounds of users on regulatory-focused subreddits, their motivations, credibility assessments, and decision-making processes, and compliance challenges. 
We followed the guidelines in \cite{prybylo2024evaluating} to minimize length and focus on users' interactions. Thus, we tailored the questions to user responses. For example, given \textit{Q17: ``When interacting on Reddit, what do you do typically? (e.g., creating the original posts, replying to other users' posts, or reading the posts)''}, we present follow-up questions based on the given response. The survey is organized into blocks as follows, using Qualtrics:

    \begin{itemize}
        \item \textit{Professional Background and Interactions:} We ask about participants' professional background (e.g., roles and company location), their privacy certifications, and the frequency and types of their engagements when seeking GDPR advice.  
         \item \textit{Motivation for Interaction:} Aligned with RQ1, we included questions that captured participants' motivations for interacting on Reddit based on their interaction types. 
         \item \textit{Perceptions and Challenges:} To address RQ2, we investigate the key obstacles participants face when implementing GDPR compliance and their perceived barriers to applying externally sourced privacy or legal advice in practice. 
         \item \textit{Credibility Evaluation and Ambiguity Resolutions Strategies:} For RQ3, we examine how users assess the credibility of advice shared and resolve unclear posts, and whether this assessment impacts their decision-making process.
    \end{itemize}
    
\noindent \textbf{Demographic Question:}
We placed questions about their demographics \textit{(Q31–Q35)}, including age, gender, education level, degree, etc., at the end of the survey to reduce early dropout and minimize bias in how participants respond to subsequent questions~\cite{fowler2013survey}. 



\subsubsection{Data Collection} 
\label{subsec: recruitment}

We recruited participants by posting a flyer on \texttt{r/gdpr} and \texttt{r/europrivacy} for one week in September 2024, and by directly messaging Reddit users. We selected \texttt{r/gdpr} and \texttt{r/europrivacy} because these subreddits have 20 times more members than smaller subreddits (e.g., \texttt{r/ccpa}) and focus specifically on regulatory compliance topics. This focus contrasts with \texttt{r/privacy}, which broadly covers privacy-related topics, and subreddits like \texttt{r/androiddev} that focus on device and platform developer challenges, such as how to request permissions. To reach active contributors and increase response rates, we also sent direct messages to selected users. These users were manually identified based on visible contributions (e.g., frequent posting, upvoted responses, and commenting) within \texttt{r/gdpr}, \texttt{r/europrivacy}, and other regulatory-focused subreddits such as \texttt{r/LegalAdviceEU}. Direct outreach was limited to users engaged in GDPR-related discussions.
These communities bring together diverse stakeholders, including developers, legal and privacy experts, and end users, whose interactions shape how compliance is interpreted and applied in practice. 

Prior to participation, users were required to agree to an IRB-approved consent form that describes eligibility, data privacy, and compensation terms. After one week, we stopped data collection and deleted the flyer from Reddit. As disclosed in the consent form, the first 100 participants, regardless of their responses, received a \$5 Amazon gift card upon completion of the $\sim$8–12 minutes survey (equivalent to $\sim$\$25/h). 
To maintain anonymity, those who wished to be compensated entered their email addresses in a separate, unlinked form at the end of the survey. Not all completed this step. 

\subsubsection{Survey Analysis Process}
\label{subsec:survey_analysis_process}
A power analysis~\cite{cohen1992statistical} indicated that approximately 100–120 total responses would provide at least 80\% power to detect medium effects (e.g., $\chi^2$ = 0.30, $\rho$ = 0.25, $f$ = 0.25). Based on this, we estimated that around 150 submissions would yield more than 80\% power. To reach this target, recruitment (described in Section~\ref{subsec: recruitment}) was conducted over one week. We removed the public survey post and stopped direct outreach after this period. 

We collected 522 submissions. Given the open recruitment strategy on Reddit and the compensation offered to the first 100 participants, we anticipated a substantial number of low-quality or automated responses. We therefore implemented a conservative multi-step validation process prioritizing data integrity over sample size. 
Instead of attention-check questions, we employed Qualtrics’ built-in tools to flag low engagement, e.g., straight-lining, bot-like speed, and duplicate entries. We also manually reviewed all 522 responses regardless of flagging for poor-quality indicators, such as short completion time (submitting \texttt{<} 1 minute), repeated text, nonsensical or blank open-ended answers, skipped or incomplete question responses, and characteristics consistent with AI-generated text ~\cite{tang2024science, munoz2024contrasting}. Submissions missing only the optional demographic questions were retained.
This process resulted in 223 valid responses, of which 80 included completed demographic information. 

In the \emph{quantitative analysis}, we used the Chi-squared test~\cite{greenwood1996guide} to determine whether there is a significant correlation between two categorical variables (e.g., whether reported GDPR challenges differed by company type, size, or location, and job role). For the Likert-scaled questions, we used the Kruskal-Wallis test~\cite{breslow1970generalized}. When both variables were ordinal, we applied Spearman’s rank correlation~\cite{spearman1961proof} to evaluate monotonic relationships. To analyze repeated-measures data, such as how responses to multiple Likert items vary by participant background, we used a Mixed ANOVA~\cite{henderson1953estimation}, treating each item as a within-subject factor and demographic variables (e.g., job role and education) as between-subject factors. We also applied a LASSO logistic regression model~\cite{tibshirani1996regression} to identify which credibility strategies best predict users’ decision-making behavior. 


We analyzed the \emph{open-ended survey questions} (e.g., privacy certifications (Q6) or steps to assess credibility (Q24)) using an open-coding procedure to identify categories based on prevalent themes in the data, until saturation was reached.  
We initially assigned codes to each response based on keywords, common terms, or themes. For example, if the response to Q24 was ``check profile'' or ``cross-fact check'', we assigned initial codes: check user profile or seek external verification.  We continued until no new codes were introduced. We then grouped similar codes to form broader categories and refined them iteratively during annotation to ensure clarity, internal coherence, and to reduce overlap. We then examined the ``Other'' code to see if a new theme emerged. After finalizing the categories, a privacy expert (i.e., the last author) reviewed the codes, categories, and responses to verify their accuracy and consistency.

\subsection{Reddit Post Analysis Approach}
\label{sec:reddit}

We cross-validate the survey findings by analyzing posts from \texttt{r/gdpr} and \texttt{r/europrivacy}. 

\subsubsection{Data Collection}
\label{subsec:reddit_post_data_collection}

We collected posts related to privacy and software development from the \texttt{r/gdpr} and \texttt{r/europrivacy} subreddits using the Python Reddit API Wrapper~\cite{praw} in April 2025. We did not filter posts based on upvotes or comment count.
Although these subreddits focus on privacy regulations, they often include content unrelated to privacy in software development, like career advice, news, surveys, or published articles. Therefore, we built a search term list similar to the Hark method~\cite{harkous2022hark}, which draws on two commonly used and complementary privacy taxonomies: Solove’s taxonomy of privacy harms~\cite{solove2005taxonomy} and the privacy-enhancing technology taxonomy~\cite{wang2009privacy}. 
Each privacy category serves as a base search term. We expanded these base terms with relevant synonyms to broaden coverage, resulting in a final set of 38 search terms (see Appendix \ref{app:privacykeywords} - Table \ref{tab:privacy_keywords} for more detail). We then queried Reddit for these terms and retrieved a total of 
2,248 posts (i.e., 2,014 from \texttt{r/gdpr} and 234 from \texttt{r/europrivacy}), where the majority of posts were created after January 1, 2020.  

\subsubsection{Ground-Truth Dataset Creation}
\label{subsec:reddit_post_labeling}
Our goal is to identify privacy challenges in software development by analyzing the extracted 2,248 posts. To achieve this, we first manually annotate a subset of posts related to both \emph{privacy} and \emph{software development} with the type of GDPR-related challenges to establish a ground truth. We then use a large language model (LLM) to automate the annotation process. Specifically, we randomly select a subset of posts from the collected Reddit data and label them in two steps: (1) determining whether each post is relevant to both privacy and software development; and (2) categorizing the relevant posts according to the GDPR-related challenges or questions they addressed. 

In Step 1, two authors with expertise in privacy and software development independently reviewed a subset of posts across multiple rounds to determine whether they were related to both \emph{privacy} and \emph{software development}. We initially randomly selected 100 posts and repeated the process until we collected at least 100 posts relevant to privacy and software development. 
In each round, we calculated the inter-rater reliability agreement statistic, Cohen's Kappa. This value ranged from 0.684 to 0.834, depending on the round, indicating ``substantial'' to ``almost perfect'' agreement. After each round, a third expert reviewed the posts with disagreements and documented their own response.
All three annotators then met to resolve disagreements before proceeding to the next round. After four rounds, we identified 116 posts out of 350 posts related to both privacy and software development (See Appendix \ref{app:labelingsoftwareposts} - Table ~\ref{tab:annotation_summary_reddit_post}).


In Step 2, the same two annotators independently reviewed these 116 posts and assigned one or more labels with a set of GDPR challenge categories, accompanied by their rationale. We followed a hybrid grounded analysis \cite{hoda2025qualitative}, where the initial categories were adopted from CIPL’s 2017 report on GDPR implementation~\cite{CIPL_GDPR_Implementation_2017}. However, we expanded the categories as new themes emerged. If the post did not match the initial challenge, the annotators first labeled them as ``Others''. They then analyzed the ``Others'' category to identify recurring themes, which led to three additional GDPR challenge labels, namely, ``consent'', ``GDPR definitions'', and ``data processing agreement (DPA)'', in Round 1. In this round, they also agreed on 79/116 posts. Thus, the annotators relabeled the entire dataset again using the updated list of challenges, achieving agreement on 98/116 posts. A third expert reviewed the disagreements and made a final decision. Appendix \ref{app:GDPRChallenegedef} - Table \ref{tab:GDPR_challenege_def} presents the initial challenges, along with those added during our analysis. 


\subsubsection{Automated Post Labeling}
\label{subsec:reddit_LLM_labels}

We automate labeling of the remaining 1,898 posts using in-context learning with LLMs~\cite{brown2020language} and a human-in-the-loop (HiTL) approach \cite{pangakis2024keeping}. 
In HiTL, the ground truth dataset is used to train and evaluate a model, which is then used to label a larger, unreviewed dataset. 
The two dataset distributions are assumed to be very similar because the ground truth data was randomly sampled from the same dataset as the larger set. If recall on the testing evaluation is high (i.e., $tp / (tp + fn) \ge 0.85$), then HiTL can reduce the investigator's effort to reject false positives without significant losses due to false negatives. For example, if ~30\% of 1,898 posts are privacy and development related, then we may miss $\sim$85 out of 569 posts as false negatives, while reliably adding 484 additional posts through manual validation.
Although unsupervised machine learning (ML) approaches, e.g., topic modeling, could be used to cluster posts and challenges, it is best applied when we do not have a priori knowledge. In this work, we instead begin with developer-reported privacy challenges from a survey and seek to identify evidence of these challenges in the Reddit posts, which requires supervised ML to classify texts into challenge categories. To that end, we supervise prompt design with demonstrations for use with in-context learning to classify the posts.

\begin{table}[t]
\centering
\caption{F1 Scores for Different Models Using CoT}
\renewcommand{\arraystretch}{1.2}
\resizebox{\columnwidth}{!}{
\begin{tabular}{|p{3.2cm}|c|c|c|}
\hline
\textbf{Model} & \textbf{Privacy} & \textbf{Development} & \textbf{Challenges} \\
\hline
Llama3.2-1B-Instruct & 0.8560 & 0.4030 & 0.3584 \\
Llama3.2-3B-Instruct & 0.8270 & 0.4211 & 0.2939 \\
Llama3.1-8B-Instruct & 0.9189 & 0.5405 & 0.4768 \\
Qwen2.5-7B-Instruct  & 0.8710 & 0.4407 & 0.6388 \\
gpt-3.5-turbo-1106   & 0.9498 & 0.6477 & 0.4019 \\
gpt-4o-mini          & \textbf{0.9535} & 0.4552 & 0.6683 \\
GPT-5-mini           & 0.9498 & 0.8008 & \textbf{0.7357} \\
GPT-5                & \textbf{0.9423} & \textbf{0.8531} & 0.6857 \\
\hline
\end{tabular}
}
\label{tab:modelcomparisontest}
\end{table}

We designed two text classification tasks: (1) separately classify posts that are related to both privacy and development, and (2) classify posts by the privacy compliance challenges raised. For both tasks, we first refined the prompts on a subset of the ground-truth dataset (posts for the development set) before testing them on the remaining subset of unseen posts (the holdout set). We also retain a subset of data for any potential prompt refinements. 


For the first classification task, we selected 100 posts ($\sim$30\%) from the ground truth as the development set and evaluated performance across both open-source and proprietary LLMs. Specifically, we tested three LLaMa, one Qwen, and four OpenAI models using three prompt engineering strategies: zero-shot prompting, 2-shot learning~\cite{brown2020language}, and chain-of-thought (CoT) prompting~\cite{wei2022chain} -- in total 24 experiments. 
Our final prompt (i.e., Appendix \ref{app:promptsengineering} - Listing \ref{listing:cot-dev-privacy}) performed consistently the best across both labels and all models. 


We then randomly sampled an additional 140 posts ($\sim$40\%) not used in the development set to test the final prompt across all models to select the best model. While GPT-4o-mini achieved the highest $F1 = 0.9535$ for the privacy label, GPT-5 produced the best overall results, with the highest $F1 = 0.8531$ for the development label and a comparable $F1 = 0.9423$ for the privacy label, which is 1.2\% below the highest score in privacy (see Table~\ref {tab:modelcomparisontest}). Thus, we adopted GPT-5 with CoT for the HiTL labeling effort.

After running our prompt on the entire dataset using GPT-5, 591 additional posts were identified as related to both privacy and software development. We manually reviewed and updated the predicted labels and removed 41 irrelevant posts. 
Overall, out of the 2,248 extracted posts, 666 posts (30\%) were identified as related to privacy and software development (550 labeled via HiTL and 116 from the ground truth). We did not review the posts labeled as unrelated. 
However, since recall was high (i.e., 0.9286 for privacy and 0.8571 for development), we likely captured most related posts.

We then applied the same process to identify the \emph{expressed challenges} (i.e., the second classification task). 
We first extended the set of 116 posts related to privacy and development by manually annotating an additional 38 posts randomly sampled from the HiTL method. This yielded 154 posts labeled with challenges for use in development and holdout sets, and to test our final prompt. The details for testing prompts and models are in Appendix \ref{app:promptsengineering}.
After running the experiments with the three prompting strategies on the same eight models, we found that the CoT prompt (see Listing \ref{listing:cot-challenge}) performs best. 
We then used our final prompt to test the models on the 62 unseen labeled posts. As shown in Table~\ref{tab:modelcomparisontest}, GPT-5-mini achieved the best performance with an $F1 = 0.7357$. Although the F1 score is not high, it is comparable to prior work for identifying privacy concepts in policy text \cite{baldwin2025prompts}. Furthermore, research \cite{Jain2022} shows that manually classifying posts is more labor-intensive and error-prone than verifying and correcting labels.
Thus, we adopted GPT-5-mini with CoT for identifying GDPR challenges in the dataset.

We then ran the final prompt on the remaining 512 posts and observed that the model assigned exactly one label per post (i.e., no post lacked a label). Next, the two annotators reviewed all posts with their assigned labels and flagged those that were incorrect. They reviewed the incorrect posts with the third privacy expert, which led to changes in the labels for 138 posts.  
Finally, the same two annotators reviewed all posts labeled as ``Others'' for potentially new themes. 
They identified two additional challenge categories that were not present when constructing the ground truth: ``data transfer'' and ``generic compliance'' (see Appendix \ref{app:GDPRChallenegedef} - Table \ref{tab:GDPR_challenege_def} for their definition).  
A senior privacy expert reviewed these final mappings to ensure their accuracy.

\begin{lstlisting}[language=HTML,  caption={CoT Prompt for Detecting GDPR Challenges},label={listing:cot-challenge}, 
basicstyle=\scriptsize\ttfamily,
breaklines=true,
]
You are an expert system that needs to determine if Reddit posts are in any way related to various GDPR challenges and to provide an explanation for your reasoning.

These are the only valid challenges and their definitions:
{labels' list}

Step 1: Review the examples of posts and their answers.
{examples' list}

Step 2: Understand how to apply this reasoning to new posts, using the examples and explanations as guidance.

Step 3: Apply this understanding to the following post.

Step 4: Remember to prioritize accuracy and clarity in your analysis, using the provided context and your expertise to guide your evaluation. If you are uncertain about the classification, choose 'Others' and explain your rationale for this uncertainty.

Step 5: Determine the correct classification for the following post. Make sure to review your response and provide a rationale for your selection:

For the following post:
- Return a valid array of containing ONLY the most applicable label from the list above that applies to the post.
- Do NOT ever include more than one label.
- Do NOT include the definition of the label in your answer, only the label name in quotes.
- If none apply, return an empty array, [].
- Do NOT include any code, only the array and your explanation.

Now label the following post.
Title: {post title}
Text: {selftext}
Explanation:
\end{lstlisting}

\section{Ethical Considerations and Limitations}


\textbf{Ethical Considerations:} The survey and post analysis were conducted under our Institutional Review Board’s review and approval. All participants were informed about the research, the risks and benefits of participation, compensation, and confidentiality before providing their consent. Participation was voluntary and withdrawable at any time. We did not collect personally identifiable information, and we have measures in place to ensure the anonymity, confidentiality, and security of responses. While contact information for all investigators and the IRB team was provided, no participants contacted them about the study or the compensation. Prior to analyzing the publicly available Reddit posts, we removed all usernames and comments and only shared post titles and body text with LLMs.



\textbf{Limitations:} We observed limitations in our research design that we attempted to mitigate.  
While we used screening questions to target individuals in privacy or software development, our sample may not fully represent all privacy practitioners or developers engaging with GDPR. Reddit users participate voluntarily, which may limit the representativeness of our sample. 
To reduce self-selection bias, we deliberately avoided advertising the survey as ``privacy research''. However, the survey remains subject to self-report bias, recall bias, and social desirability bias — e.g., by overreporting certifications or involvement in GDPR compliance. Despite efforts to validate submissions, including the use of Qualtrics' bot detection and manual screening,  some responses may have been generated or influenced by AI tools. We conservatively excluded suspicious entries, which may have inadvertently removed legitimate responses. 

We used a 4-point Likert scale to reduce response bias, including central tendency (avoiding clear choices) and social desirability bias (giving socially acceptable answers), as respondents often select the midpoint when they are unsure or wish to avoid disagreement~\cite{chyung2017evidence, garland1991mid}. 
Our survey used individual agreement-based items rather than a composite Likert scale that aggregates multiple statements measuring a single construct. This approach allowed us to capture perceptions on specific topics while minimizing bias. 

We allowed skipping questions and avoided emotionally charged wording to minimize recall bias.
Some items, e.g., \textit{Q10: ``Within my company, the knowledge about privacy compliance is up to date''}, may be perceived as leading but were intended only to assess organizational awareness rather than individual knowledge. To strengthen our findings' validity, we conducted statistical analyses to identify significant relationships. Nonetheless, these limitations may affect results' accuracy; we thus advise caution in interpreting them.



We used predefined challenge categories from the 2017 CIPL GDPR implementation report~\cite{CIPL_GDPR_Implementation_2017} in our survey, as well as an open-ended option to capture newly emerging categories. However, requiring participants to type new challenges may have led to inconsistent capture of new categories, unlike in the post analysis. 

While our findings highlight patterns in two regulatory-focused subreddits, they may not be generalizable to other communities or regulatory contexts, such as the CCPA, which future work could address.  
Although our keyword-based search approach used terms from comprehensive privacy taxonomies, some relevant posts may have been missed. Since these terms and their synonyms cover the major privacy themes, most relevant posts were likely captured.


\begin{table*}[h]
    \centering
    \caption{Professional Background of the Participants}
    \footnotesize
    \renewcommand{\arraystretch}{1.2}
    \resizebox{\textwidth}{!}{
        \begin{tabular}{|l|l|}
            \hline
            \textbf{Category} & \textbf{Details} \\ 
            \hline
            \textbf{Company Type} & For-profit (51\%), Non-profit (10\%), Educational (15\%), Government (8\%), Research (9\%), Self-employed (6\%), Other (1\%) \\ 
            \hline
            \textbf{Company Location} & North America (50\%), Europe (14\%), Central America (13\%), South America (13\%), Africa (6\%), Other (4\%) \\ 
            \hline
            \textbf{Years at Current Job} & 0-2 Years (7\%), 3-5 Years (45\%), 6-10 Years (41\%), 11+ Years (8\%) \\ 
            \hline
            \textbf{Years in Privacy/Data Roles} & No experience (4\%), 1-2 Years (21\%), 3-5 Years (50\%), 6-10 Years (23\%), 11+ Years (3\%) \\ 
            \hline
            \textbf{GDPR Involvement} & Yes (80\%), No (14\%), Unsure (4\%), Prefer not to say (2\%) \\ 
            \hline
        \end{tabular}
        }
\label{table:professional_background}
\end{table*}

\section{Survey Findings}
\label{sec:findings}

In this section, we report our findings related to RQ1-RQ3.

\subsection{Demographics, Professional Background, and Reddit Interaction Patterns}

Although we had 223 valid participants, we received only 80 responses to the optional demographic questions (a 64\% non-response rate) as they were placed at the end of the survey. Appendix~\ref{app:demo} - Table \ref{table:demographics2} presents the demographic findings: the majority of respondents identified as male (73\%, n = 58) and are mostly between the ages of 35-44 and 25-34, representing 46\% (n = 37) and 40\% (n = 32) of respondents, respectively. 
Professional degrees (i.e., certificate degrees) are held by 39\% (n = 31) of participants, followed by 34\% (n = 27) who have completed graduate studies. The majority hold degrees in computer science (35\%, n = 28), data science (23\%, n = 18), and information technology (21\%, n = 17). 30\% (n = 24) work in companies with 101+ employees, while 51\% (n = 41) work in medium-sized companies (21-100 employees).

In contrast, for professional background questions (i.e, \textit{Q1-Q8}), we received 223 responses, as shown in Table \ref{table:professional_background}. Most participants work in for-profit companies, half are based in North America (although the subreddits primarily discuss EU laws), and have 3–5 years of job experience. About 80\% report involvement in GDPR-related tasks or compliance. Among the roles, Technical Roles have the highest representation (48\%), followed by Management and Executive Roles (23\%), Others (16\%), Academic and Research (9\%), and Legal and Compliance (4\%) (see Appendix~\ref{app:jobrole} - Figure~\ref{fig:Annotated job title}). We further break technical roles into more granular categories. As shown in Appendix~\ref{app:jobrole} - Figure~\ref{fig:Technical Roles subcategories}, within the technical roles group,  40\% are in ``software development'', 32\% in ``data science'', and 14\% each in ``cybersecurity'' and  ``IT operations'' roles.





\begin{figure}[h]
     \centering
        \includegraphics[width=0.38\textwidth]{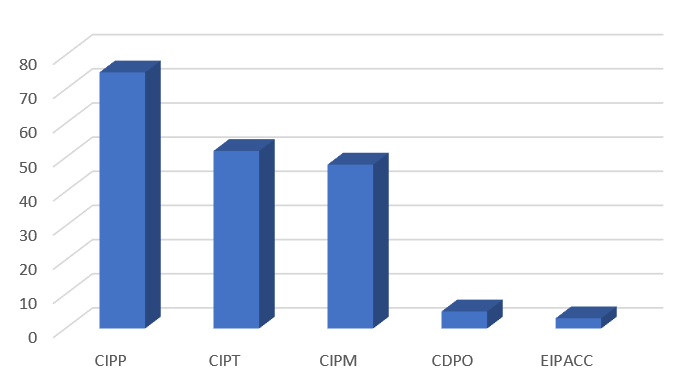} 
        \caption{Top Five Privacy Certifications Categories}
        \label{fig:Q6 Subcategories}
\end{figure}

We evaluated whether participants held any professional certifications. 
We categorized the 195 valid responses into six groups. 
Privacy-related certifications dominate the responses, with $\sim$90\%, whereas the other categories were mentioned only a few times (See Appendix~\ref{app:cert} - Figure~\ref{fig:Q6 Responses}). We further analyzed the privacy certifications in more detail. Participants listed certifications such as CIPP (41\%), CIPT (28\%), CIPM (26\%), CDPO (3\%), and EIPACC (2\%) (see Figure~\ref{fig:Q6 Subcategories}). Most certifications are issued by the International Association of Privacy Professionals (IAPP)~\cite{IAPP-Certifications}, which is particularly popular in North America. EU-specific certifications such as CDPO~\cite{CDPO} and EIPACC~\cite{EIPACC} were less frequently mentioned, likely reflecting the geographic distribution of respondents. 

\begin{figure*}
    \centering
    \begin{subfigure}{0.3\textwidth}
        \centering
        \includegraphics[width=\textwidth]{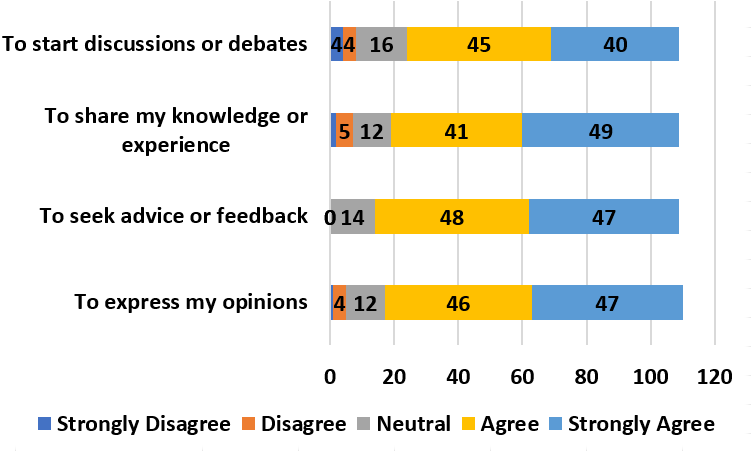} 
        \caption{Motivations for Create Posts (Q19)}
        \label{fig:Q19}
    \end{subfigure}
    \hfill
    \begin{subfigure}{0.3\textwidth}
        \centering
        \includegraphics[width=\textwidth]{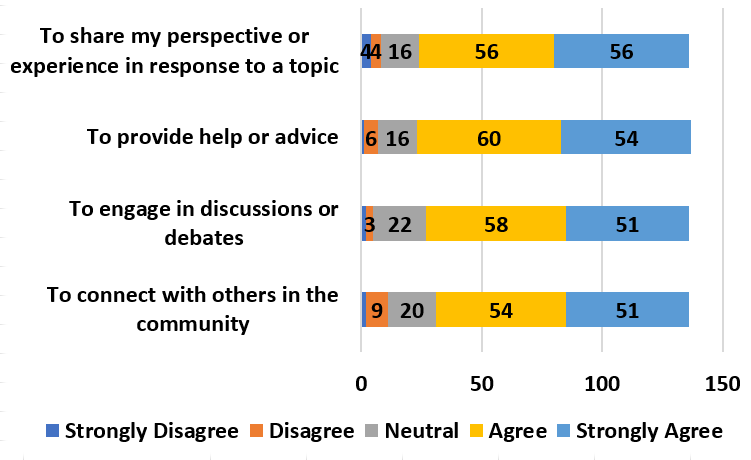} 
        \caption{Motivations for Post Replies (Q21)}
        \label{fig:Q21}
    \end{subfigure}
    \hfill
    \begin{subfigure}{0.3\textwidth}
        \centering
        \includegraphics[width=\textwidth]{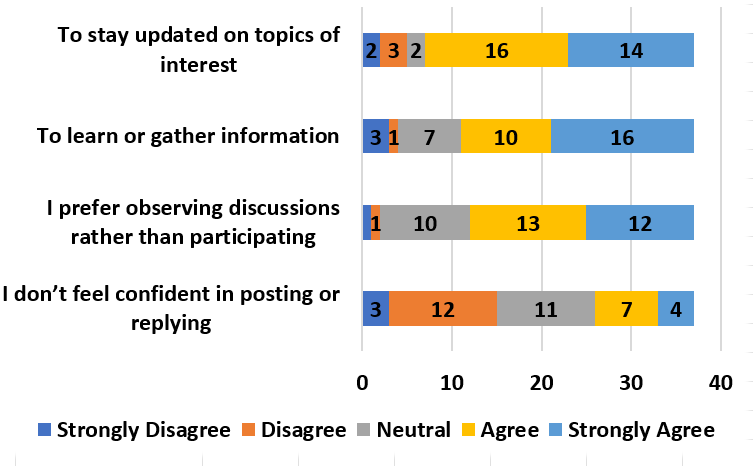} 
        \caption{Motivations for Only Read (Q23)}
        \label{fig:Q23}
    \end{subfigure}
    \caption{Motivations For Interaction on Reddit}
\end{figure*}


We explore whether participants interact with Reddit and, if so, what the nature of their interaction is. As shown in Appendix~\ref{app:Frequencyfig58} - Figure~\ref{fig:58}, out of 223 respondents, $\sim$75\% reported that they engage with GDPR-related posts at least a few times a week or daily, similar to the findings in \cite{prybylo2024evaluating}. $\sim$8\% of users almost always post or reply about GDPR topics, while about 50\% have posted or replied about GDPR, frequently in the last six months (see Appendix~\ref{app:Frequencysixmonths} - Figure~\ref{fig:Fr6months}). The most common behavior among participants is to equally post and reply to others (34\%), followed by mostly replying to others’ posts (31\%) and primarily creating posts (19\%). About 17\% mostly read posts without posting or replying (see Appendix~\ref{app:Frequencys} - Table~\ref{tab:reddit_interaction}). 




\subsection{Users' Motivation for Interaction}

To address \emph{RQ1}, we explore the motivations for engaging in discussions on the subreddits and examine whether these motivations relate to users’ professional, organizational, and privacy-related backgrounds. 
Depending on the interaction types (see Table~\ref{tab:reddit_interaction}), we asked tailored follow-up questions \textit{(Q18–Q23)}. 

\textit{Users' Motivations for Creating Original Posts:} 
78 participants described their motivation for creating original posts on GDPR-related subreddits (including those who equally create or reply). 
We noticed that most mentioned they post to share their knowledge or experience (37 responses, 47\%), express opinions (19 responses, 24\%), seek advice or feedback (10 responses, 13\%), and start discussions or debates (8 responses, 10\%).  
For example, one respondent said 
\emph{“I create posts on the importance of security”}. Several noted that \emph{``they post when they consult external sources for privacy or legal compliance and want to share those findings with the community''}. About 5\% are labeled as ``Other'', as they do not fit into any categories.

We then asked them to rate their motivations on a Likert scale to validate the open-ended responses. We received 110 responses.  
83\% agreed or strongly agreed that they are motivated by a desire to share their knowledge or experience (see Figure \ref{fig:Q19}). Seeking advice or feedback and expressing their opinions are also key motivators for 86\%  and 85\% of respondents (at least agreeing), respectively. Finally, 78\% agreed or strongly agreed that starting discussions or debates encourages them to share original content. These findings are aligned with the open-ended responses. 

\textit{Users' Motivations for Post Replies:} 
87 participants expressed their motivation for replying to others' posts. 
Most replied to provide help or advice (39 responses, 45\%), share their perspective or experience (27 responses, 31\%), engage in discussions or debates (9 responses, 10\%), or connect with others in the community (5\%). For example, one reported 
\emph{“providing help and advice, sometimes engaging in discussion”}, while some said they specifically aim to challenge or validate claims, e.g., \emph{“cross-check with multiple sources before replying”}. 9\% were labeled as ``Other'', (e.g., motivations like 
\emph{``telling others how impressed I am''}) which did not fit our themes.

138 participants ranked their motivation on a Likert scale. 
The majority (83\%) agreed that they are motivated to provide help or advice. They also (strongly) agreed with: sharing their perspectives or experiences (82\%); connecting with others in the community (77\%); and engaging in debate (80\%) (see Figure \ref{fig:Q21}).

\textit{Users' Motivations for Reading without Engaging:} 
We received 19 responses from those who mostly read the posts without posting or replying. 
Nearly half (9 responses, 47\%) reported reading primarily to learn or gather information. Other motivations include staying updated on topics of interest (3 responses, 16\%), preferring to observe discussions rather than actively participate (3 responses, 16\%), and lacking confidence in posting or replying (2 responses, 11\%). Two responses were labeled as ``Other.'' Examples include \emph{“I read and analyze to stay up to date with emerging trends”} and 
\emph{“learn about it first and then share it with people around you”}.

A total of 37 participants responded to the Likert-scale prompt. As shown in Figure \ref{fig:Q23}, the most common motivations are staying updated on topics of interest (81\%), learning or gathering information (70\%), and preferring to observe discussions rather than participate (68\%). A smaller proportion (30\%) indicated they read posts because they do not feel confident in posting or replying.

We also examined whether the types of motivations relate to users’ job roles \textsf{(H1a)}, privacy-related
backgrounds, i.e., privacy-related certifications \textsf{(H1b)}, and privacy-related experience \textsf{(H1c)}, and company size \textsf{(H1d)}. We did not observe any statistically significant relationships between job roles (N = 27–118, depending on the motivation) and their motivations. To evaluate \textsf{(H1b)}, we grouped participants into two categories: those with privacy-related certifications and those without. Participants holding \emph{privacy-related certifications} 
reported significantly stronger motivations for creating posts to share knowledge and experiences ($U = 956.50$, $p-value = 0.001$, $r = 0.476$, $N = 94$) and to express opinions ($U = 859.50$, $p-value = 0.032$, $r = 0.310$, $N = 95$) compared to others. We also observed significant relationships between \emph{privacy-related experience} and motivations for creating posts to share knowledge and experiences ($H = 14.30$, $p-value = 0.006$, $\epsilon^2  = 0.112$, $N = 95$) and replying to share perspectives or experiences ($H = 8.80$, $p-value = 0.032$, $\epsilon^2  = 0.051$, $N = 117$). Company size was associated with motivations for seeking advice and feedback ($H = 10.96$, $p-value = 0.027$, $\epsilon^2  = 0.279$, $N = 30$) when creating posts and replying to share perspectives or experiences ($H = 9.02$, $p-value = 0.029$, $\epsilon^2  = 0.140$, $N = 47$). These findings suggest that privacy expertise and organizational context may influence motivations for participating in GDPR-related online discussions. 

\begin{tcolorbox}[colback=gray!10,colframe=black!30,boxrule=0.1pt, boxsep=0.1pt]
\textbf{Summary:} Regulatory-focused subreddits support both advice-seeking and knowledge-sharing around compliance. Privacy-related experience, certifications, and organizational context may shape how users participate in these discussions. 
\end{tcolorbox}

\subsection{Perceptions and Challenges}
\label{tab:perception-challenges}

We examine GDPR challenges and perceptions of compliance across professional and organizational backgrounds (i.e., \emph{RQ2}).

We asked participants to describe the aspect of GDPR compliance they find most challenging, with response options adopted from CIPL’s 2017 report on GDPR implementation~\cite{CIPL_GDPR_Implementation_2017}. We received 220 responses. Multiple options were allowed (see Table~\ref{tab:perception-challenges}). 
121 respondents (55\%) considered data protection impact assessments (DPIAs) as the most challenging. Data breach notification is the second-most reported challenge (49\%, 107 responses), followed by data subject access requests (DSARs) (39\%, 85 responses), and maintaining records of processing activities (RoPA) (29\%, 64 responses). Only 1\% (3 responses) selected ``Other, please specify'' which included \emph{``data retention issues''},  \emph{``designing compliant personal data flows across cloud systems and organizations''}, and \emph{``the whole process and divergence of opinions making arbitrations difficult''}. The results suggest that the processes for minimizing risk are perceived as the most challenging aspect of GDPR compliance. 

We also explore whether participants’ company characteristics, i.e., type (\textsf{H2a}), size (\textsf{H2b}), and location (\textsf{H2c}), job roles (\textsf{H2d}), as well as holding privacy-related certifications \textsf{(H2e)}, influence their perception of these challenges (See Appendix \ref{app:Hi2list}). While we find significant relationships between company type or size, and holding privacy-related certifications and GDPR challenge selection ($N = 190$) with ($p-value = 0.0001$, $Cramér’s V = 0.2123$), ($p-value = 0.0195$, $Cramér’s V = 0.1564$), and ($p-value < 0.001$, $Cramér’s V = 0.2819$), the company's location or job roles show no such relationship 
(see Appendix~\ref{app:Hi2list} - Table \ref{tab:h2results}). 
These results indicate that although companies of different types and sizes focus on various compliance issues and face distinct regulatory concerns, within the company, the focus on compliance issues is homogeneous across management, legal, and development teams, except for those holding privacy-related certifications. For example, for-profit companies often selected challenges such as data breach notification and DPIAs, while government and academic groups emphasized these relatively less and highlighted other concerns. Whether the company is located in North America or the EU does not have an impact, likely because the focus is only on GDPR challenges.  


\begin{table}[h]
\centering
\caption{The Distribution of GDPR Challenges}
\label{tab:challenges}
\begin{tabular}{|p{5.8cm}|c|}
\hline
\textbf{Category} & \textbf{\% (Count)} \\
\hline
Data Protection Impact Assessment (DPIA) & 55\% (121) \\
\hline
Data Breach Notification & 49\% (107) \\
\hline
Data Subject Access Requests (DSARs) & 39\% (85) \\
\hline
Records of Processing Activities (RoPA) & 29\% (64) \\
\hline
Other & 1\% (3) \\
\hline
\end{tabular}
\end{table}

Next, we check whether the participants' organization has any in-house legal team specializing in privacy compliance. The majority (62\%, 137 respondents) indicated that their organization has a dedicated privacy or legal team. An additional 25\% (56 respondents) reported having a legal team that is not explicitly dedicated to privacy. However, 10\% noted the absence of an in-house legal team, while 2\% were unsure. These findings show that while 87\% of organizations have some form of in-house legal support, only 62\% have teams dedicated explicitly to privacy compliance. This limited in-house dedicated privacy teams suggests that many organizations may need supplementary support. This is further reflected in participants’ reports of whether they or their companies seek external privacy or legal advice. As shown in Appendix~\ref{app:q12} - Figure \ref{fig:Q12}, only 13\% (29 respondents) said they manage all compliance internally, while 9\% (21 respondents) rarely seek outside help, and 5\% (10 respondents) were unsure. The majority (73\% of participants) report seeking external privacy or legal advice at least occasionally. We also notice that those in small companies (i.e., $<100$) reported higher reliance on external legal or privacy advice (i.e., 80\%). 
These results indicate that companies routinely compensate for gaps in internal privacy expertise by engaging external specialists or other tools, highlighting an underlying mismatch between regulatory demands and available in-house capabilities. The results are also aligned with findings in 
\cite{prybylo2024evaluating} where a privacy officer role was not often present, even in larger companies, and  $\sim$50\% seek external advice, check the best practices like NIST guidelines, or use forums. 

We assessed whether participants perceive their company's privacy compliance knowledge to be up to date. A total of 220 responded. Most expressed confidence in their organization's privacy compliance knowledge, with 86\% selecting ``agree'' or ``strongly agree''. 10\%, were neutral, while only 3\% (strongly) disagreed.  

We then evaluated whether their confidence in their company’s privacy compliance is influenced by the presence of internal legal resources (\textsf{H3a}) and reliance on external legal advice (\textsf{H3b}) (See Appendix \ref{app:Hi3list} - Table \ref{table:H2l_pvalues}).  
With $H = 8.4773$, $p-value = 0.0371$, $\eta^2 = 0.0294$, $N = 190$, we find a significant relationship for \textsf{H3a}, but 
did not observe a statistically significant relationship for \textsf{H3b}, 
with $\rho = 0.0083$, $p-value = 0.9096$, $N = 191$, highlighting that a dedicated privacy or legal team increases the likelihood that employees perceive their company’s privacy knowledge as up to date (similar to \cite{prybylo2024evaluating}), but this confidence is largely consistent, regardless of whether and how often they relied on outside legal support. 

We received 221 responses regarding their primary sources of information for staying updated on GDPR. As shown in Table \ref{tab:gdpr_sources}, the most common sources are official regulatory websites (57\%, 126 respondents) and professional networks and forums (51\%, 112 respondents), suggesting that they prioritize authoritative resources and peer communities. Industry conferences and webinars (43\%, 95 respondents) and legal and compliance publications (36\%, 79 respondents) are also considered. Only 24\% cited internal corporate training, indicating its limited role in ongoing education, aligned with findings in \cite{horstmann2025searchingprivacy}. 3 respondents selected ``Other.'' We examined whether primary sources of GDPR knowledge vary based on their job roles (\textsf{H4a}), company size (\textsf{H4b}), educational background (\textsf{H4c}), and privacy-related certifications \textsf{(H4d)} (See Appendix \ref{app:Hi4list} - Table \ref{table:H3_pvalues}). 
We do not find statistically significant relationships between GDPR knowledge sources and job roles, 
company size, 
or educational background. 
However, we observed a significant relationship for privacy-related certifications \textsf{(H4d)}, with $p-value = 0.0046$, $Cramér’s V = 0.2820$, $N = 191$. These findings suggest that the sources of GDPR information do not differ substantially across professional, company, or educational backgrounds. However, users with privacy-related certifications may rely on different sources. 

\begin{table}[t]
    \centering
    \caption{Primary GDPR Information Sources}
    \begin{tabular}{|p{5.5cm}|c|}
        \hline
        \textbf{Source of Information} & \textbf{\% (Count)} \\
        \hline
        Official regulatory websites (e.g., ICO) & 57\% (126) \\
        \hline
        Professional networks and forums & 51\% (112) \\
        \hline
        Industry conferences and webinars & 43\% (95) \\
        \hline
        Legal and compliance publications & 36\% (79) \\
        \hline
        Internal corporate training & 24\% (53) \\
        \hline
        Other & 1\% (3) \\
        \hline
    \end{tabular}
    \label{tab:gdpr_sources}
\end{table}




\begin{tcolorbox}[colback=gray!10,colframe=black!30,boxrule=0.1pt, boxsep=0.1pt]
\textbf{Summary: }Participants consider DPIAs and data breach notifications the most challenging aspects of GDPR compliance. Challenges vary by company type and size, and holding a privacy-related certification. Most companies rely on external legal help to fill gaps in their internal privacy expertise.
\end{tcolorbox}

\subsection{Credibility Evaluation \& Ambiguity Resolutions Strategies}

We answer \emph{RQ3} by exploring how participants evaluate the credibility of replies or how they disambiguate before responding.

\begin{figure}
     \centering
        {\includegraphics[width=0.42\textwidth]{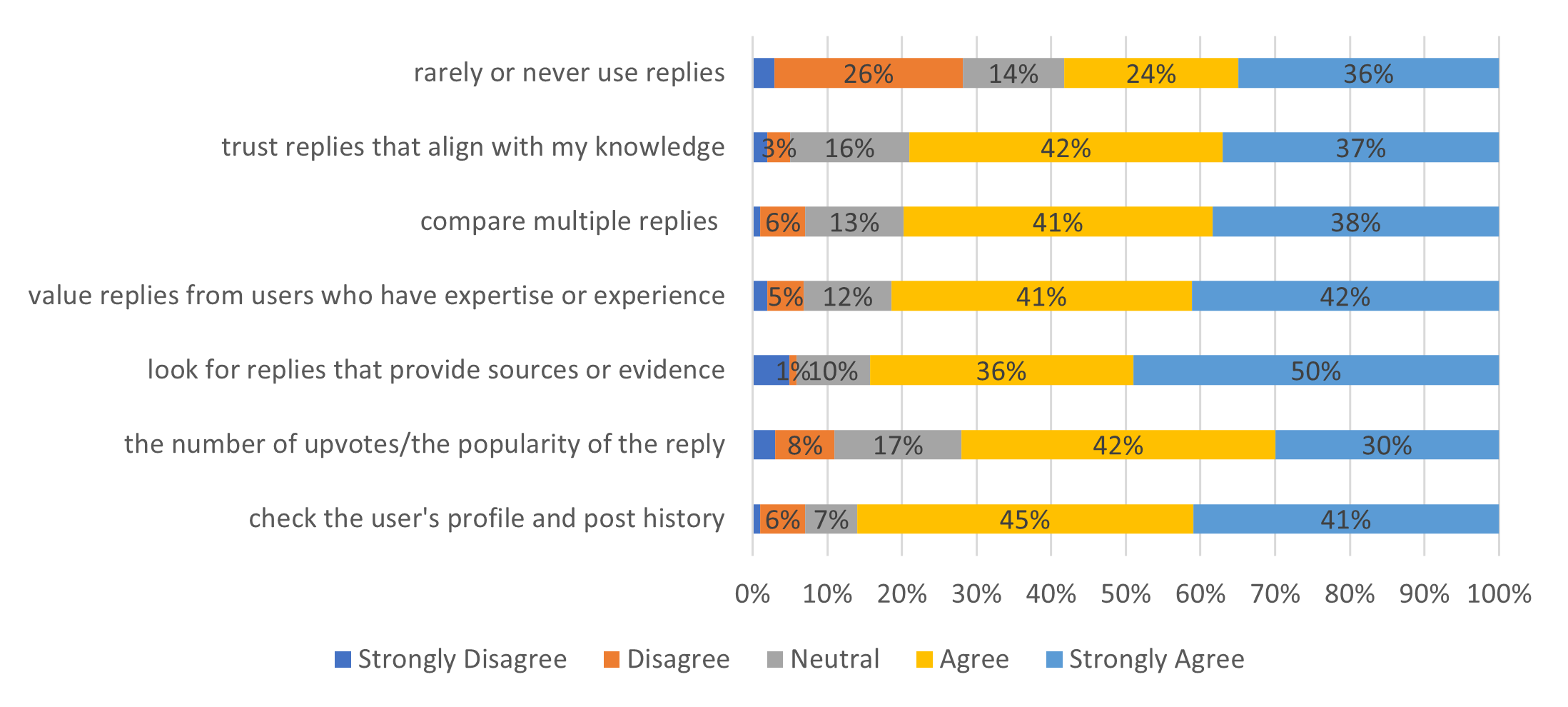}} 
    \caption{Credibility Evalution Strategies}
    \label{fig:(Q25)}
\end{figure}

Participants who created posts were asked how they decide whether the reply is credible. We manually categorized the open-ended responses into eight groups, 
as shown in Table~\ref{tab:info_evaluation_methods}. For example, 22\% mention they check \textit{the user profile and history}, e.g., ``verify profile and previous posts including replies.''
18\% of them seek \textit{external verification} including ``Google and fact checking website'' and ``reputable news articles across vast news outlets'', while 17\% look for \textit{sources/evidence} through strategies such as ``check original post, look for the source citation'' or ``look for supporting evidence in the response.'' Smaller portions (8\%, 7\% and 6\%, respectively) search for \textit{community feedback}, e.g., ``review comments'' and ``observe feedback from other users,'' just use the \textit{responses directly} e.g., ``I use them to make decisions directly related to the topic,'' or focus on 
\textit{critical thinking or logic}, like ``I take a minute to understand what the topic is all about.'' Some participants use \emph{multiple tactics}, such as ``evaluate subreddit and poster's history, check for verified badges, look for sources, search for additional evidence, and crosscheck externally.'' $\sim$11\% do \emph{minimal verification} or \emph{were unsure}. Figure \ref{fig:(Q25)} shows the participants' rating of their agreement with seven strategies for credibility assessment. Most agreed with relying on replies that include evidence (86\%) or checking user profiles and history (86\%) while other strategies were less commonly agreed on.
Overall, users prioritized expertise and users' history over popularity, but a small number consider minimal verification.

We used mixed ANOVA analysis to examine whether participants from different professional backgrounds, including job role (\textsf{H5a}), educational background (\textsf{H5b}), company size (\textsf{H5c}), and privacy-related certifications \textsf{(H5d)}, differed in how they evaluated the credibility of replies and whether they preferred different credibility evaluation strategies. Detailed statistical results are provided in Appendix~\ref{app:H6list} - Table~\ref{table:Q25_anova_pvalues}. 
The between-subject groups are participants’ job roles, company sizes, educational backgrounds, and privacy-related certification status, while the within-subject factor is the set of credibility evaluation strategies. A main effect of strategy indicates that participants generally preferred some credibility evaluation strategies over others. An interaction effect indicates that different groups preferred different credibility evaluation strategies.
For job role ($N = 94$), we found a significant main effect ($p-value=0.0498$), but no interaction ($p-value=0.127$), suggesting that while roles differ in overall agreement, they rank the strategies similarly. For education ($N = 29$), neither the main effect ($p-value=0.719$) nor the interaction ($p-value=0.925$) was significant, indicating little influence on strategy use. For company size ($N = 29$), the main effect was not significant ($p-value=0.126$), but we observed a significant interaction ($p-value<.001$), indicating that strategy preferences varied by company size. For privacy-related certifications ($N = 94$), we did not observe a significant main effect ($p-value=0.452$) or interaction between certification status and strategy ($p-value=0.569$), indicating that participants with and without privacy-related certifications ranked credibility evaluation strategies similarly. Overall, while privacy, professional, and organizational factors may influence credibility evaluations, most users rely on similar strategies regardless of background.




\begin{table}[t]
    \centering
    \caption{Strategies for Verifying Responses}
    \begin{tabular}{| p{5cm} |c|}
        \hline
        \textbf{Category} & \textbf{\% (Count)} \\
        \hline
        Check user profile/history & 22\% (15) \\ \hline
        Seek external verification & 18\% (13) \\ \hline
        Look for sources/evidence & 17\% (12) \\ \hline
        Community feedback & 8\% (6) \\ \hline
        Use response directly & 7\% (5) \\ \hline
        Critical thinking or logic reasoning & 6\% (4) \\ \hline
        Comprehensive strategy (multi-step) & 11\% (8) \\ 
        \hline
        Minimal / Don’t know / Unclear & 11\% (8) \\ \hline
        \textbf{Total} & 71 \\
        \hline
    \end{tabular}
    \label{tab:info_evaluation_methods}
\end{table}


Participants who mainly replied to others were asked how they respond when encountering an unclear or incomplete post. Their open-ended responses showed various strategies. $\sim$50\% mentioned \emph{seeking clarification}. E.g., one participant reports that \emph{``politely ask clarifying questions, provide construction feedback, and share relevant insights''} while another said \emph{``request for a complete post from the users post''}. $\sim$13\% reported they \emph{avoid replying to unclear posts} where one said \emph{``I don't respond to unclear post that I haven't understood''}. A small group used other strategies like  \emph{``sending a private message to the subreddit''} or \emph{``say so and elaborate on my interpretation of the question.''} 
Overall, users relied on clarification, interpretation, and flexible responses when dealing with ambiguous posts.

We used LASSO logistic regression \cite{tibshirani1996regression} to determine whether credibility assessment strategies can predict how users decide on the responses. For “those who use the responses directly to make decisions”, the model selected three strategies: checking upvotes, checking the user’s profile and history, and trusting replies that match the user’s existing knowledge, with coefficient values of 0.422, 0.385, and 0.049. These users appear less focused on validating sources and tend to be less critical. 
The model did not select any strategy as a predictor for “those who use the responses as a reference, but verify the information independently", suggesting that independent verification is not explained by credibility strategies in our survey. The model selected all seven strategies for “the users who use the response as part of many opinions to form their view", but none of the predictors showed a significant effect, which indicates they rely on many small strategies together, rather than one or two major ones. For “those who mainly ignore responses unless they are highly voted or come from a credible source", two predictors were significant (i.e., pay attention to upvotes or rarely use responses with coefficient values of 0.791 and 0.641). 
Lastly, for “those who only want to begin a discussion or ask follow-up questions", the model selected four predictors with coefficient values of 0.700, 0.537, 0.498, and 0.025, where three are statistically significant. These findings suggest that users rely on different credibility strategies depending on how they use Reddit replies. 



\begin{tcolorbox}[colback=gray!10,colframe=black!30,boxrule=0.1pt, boxsep=0.1pt]
\textbf{Summary:} Most users assess credibility using evidence-based methods like verifying external sources, checking profile histories, or upvotes, and seek further clarification for ambiguous posts; yet, a smaller number never conduct verification or look for sources. Selecting one strategy over another is explained by some professional backgrounds, but privacy experience does not affect such decisions. 
\end{tcolorbox}

\section{Reddit Posts Analysis Results}
\label{sec:reddit_post_findings}

In this section, we answer \emph{RQ4}. The Reddit post analysis described in Section~\ref{sec:study} mapped each post to one or more GDPR challenge categories. Table~\ref{tab:reddit_post_challenges} shows the distribution. Among the 666 posts, consent was the challenge most frequently discussed (38\%), followed by data processing agreements (DPAs) (12\%) and data transfer to a third country (12\%). References to GDPR definitions, DPIAs, and DSARs also appeared regularly. Less frequent challenges included generic compliance, RoPA, and data breach notifications.

\begin{table}[t]
\centering
\caption{GDPR Challenges in Annotated Reddit Posts}
\label{tab:reddit_post_challenges}
{\footnotesize $^{*}$ Indicates GDPR challenges also identified in the survey.}
\vspace{1mm}
\begin{tabular}{|p{5.8cm}|c|}
\hline
\textbf{Category} & \textbf{\% (Count)} \\
\hline
Cookie Consent& 20\% (132) \\
\hline
General Consent& 18\% (119) \\
\hline
Data Processing Agreements (DPAs) & 12\% (81) \\
\hline
Data Transfer & 12\% (77) \\
\hline
Other & 9\% (62) \\
\hline
GDPR Definitions & 8\% (55) \\
\hline
\textit{Data Protection Impact Assessment (DPIA)$^{*}$} & 6\% (39) \\
\hline
\textit{Data Subject Access Requests (DSARs)$^{*}$} & 6\% (38) \\
\hline
Generic Compliance & 5\% (32) \\
\hline
\textit{Records of Processing Activities (RoPA)$^{*}$} & 4\% (24) \\
\hline
\textit{Data Breach Notification$^{*}$} & 1\% (10) \\
\hline
\textbf{Total GDPR Challenges} & 100\% (669) \\
\hline
\end{tabular}
\end{table}

\subsection{Reddit Post Analysis vs. Survey Findings}

\subsubsection{GDPR Challenge Overview} Our Reddit post analysis complements and extends the survey findings, and also reveals notable differences in concerns. While DPIAs (55\%) and data breach notification (49\%) are the most frequently cited challenges in the \emph{survey} (see Table \ref{tab:challenges}), Reddit discussions focused more on consent (38\%), particularly cookie consent, which accounted for $\sim$53\% of all consent-related posts. This difference is mainly due to the broader scope and open-ended nature of Reddit discussions, reflected in the high proportion of posts labeled as ``Others'' (9\%) and six additional categories absent in the survey. The challenges that overlap are shown in \emph{italics} with $^{*}$. As shown in Table~\ref{tab:reddit_post_challenges}, the first overlapping challenge, i.e., DPIA, is the seventh most common challenge in the post analysis. Our analysis reveals that only 111 of the 666 posts (17\%) matched challenge categories explicitly included in the survey. 
This result confirms that the survey captured only a subset of the real-world GDPR challenge space and that practitioners encounter a wider range of issues in practice than those reported by CIPL~\cite{CIPL_GDPR_Implementation_2017}. %
When we filter posts to include only those labeled with survey categories (Appendix~\ref{app:shared_gdpr_challenges} - Figure~\ref{fig:reddit_analysis_label_exist_dist}), the distribution of challenges mostly matches, with DPIAs, DSARs, and RoPA appearing in the same relative order across both datasets. The only exception is data breach notifications, which are rarely mentioned in the posts, likely due to the risks of publicly disclosing such incidents~\cite{ford2021impact, ford2023impact, spanos2016impact}.

\subsubsection{Interaction Types}
We analyzed whether posts in each category contained follow-up responses from the poster. Our analysis showed that 83.33\% of RoPA posts had follow-up comments from the original poster, followed by 78.18\% of GDPR definitions, 75\% of generic compliance, and 70\% of breach notification posts. These comments mostly covered questions and clarifications, suggesting the complexity of RoPA challenges.

\subsubsection{Analysis of Regulatory Sources} 
We also evaluated which official documents or GDPR articles were most cited in responses. ICO and EDPB were cited the most (i.e., 165 and 118 times), while IAPP and NIST sources were rare (unlike the survey findings, where NIST was more commonly used). As for GDPR, Art. 28 (Processor) has been cited the most, followed by Art. 6 (Lawfulness of processing), Art. 13 (Information to be provided where personal data are collected from the data subject), and Art. 3 (Territorial scope). Articles less frequently discussed include Art. 30 (RoPA), Art. 44 (General principle of transfers), and Art 46. (Transfers subject to appropriate safeguards); even though 16\% of challenges were tied to them. Art. 35 (DPIA) was only directly referenced once, potentially due to users using the term ``DPIA'' rather than the article number. We identified 582 (11.3\% of all comments) total references to specific GDPR articles within the comments in our dataset.

\subsection{Analysis of GDPR Challenges}
\subsubsection{Cookie Consent} Most posts in this category are concerned about what can be tracked without asking for consent, the difference between strictly necessary cookies and other types, and confusion about GDPR vs. the ePrivacy Directive (ePD). A poster asked \emph{``if consent was required for cookies implementing first-party analytics.''}

A large portion of the posts discuss Google Analytics (GA) and the consent requirements for using it. One poster sought \emph{``May I store GA cookies without consent under the (ePD)?''} Some of the posts also asked whether a personal blog requires cookie banners. Documenting and storing consent for audit purposes was another recurring question. The responses mainly focused on whether tracking is strictly necessary or not. However, there was occasional confusion regarding ``strictly necessary'' vs. ``functional'' cookies. While some users claim that consent is not required for functional cookies, one responder mentions \emph{``The author of the page you're citing has invented the distinction between strictly necessary and functional cookies out of thin air. [..] this distinction has no basis in the actual laws or in authoritative guidance from data protection agencies.''} A few posts discussed ``if a decline button is required,'' and the use of generators for cookie consents. ePD Art. 5 and GDPR Art. 7 were mentioned 16 and 14 times, respectively, while the EDPB's requirements for cookie consent were recommended only twice.

\subsubsection{General Consent} Most posts in this category involve questions about when to ask for consent, what constitutes ``lawful processing," and what falls under ``legitimate interests''. A poster asked about \emph{``..a small business that has an application to save in-store credit for their clients. The only data being stored is the client's first and last name and how much in-store credit they have. [..] Do I need some written consent [..] to store their name?''} The responses were mixed and sometimes conflicting. Some argued that managing in-store credit is a ``legitimate interests'', while others noted that GDPR would apply as they ``collect personal data''. Other posts focused on when to ask explicit or implicit consent, what information should be included, and how to carry out age restrictions or obtain parental consent. A poster sought \emph{``[..] our application is not DIRECTLY targeted to users <13 or <16, what would be the best action to go around this? Restrict users <13 [..]? and have <16 users requested parental consent?"} The responders mostly cited ``The UK Information Commissioner's Office (ICO)'' documentation or GDPR Art. 8. 

Some posts were concerned about what to include in privacy policies, e.g., \emph{``list all the 8 rights in a Privacy Policy''} or \emph{``Does GDPR REQUIRES that companies publish detailed list (name, activities, location, etc) of all sub processors the company uses?''} Similarly, the responses were often contradictory: while the majority mentioned that they ``do not have to list it *publicly* in the Privacy Policy'', a few suggested to ``give a list of where data will be shared''. In these cases, it is up to the poster to follow up, refer to GDPR articles or guidelines, etc. to verify the responses. GDPR Art. 4, 6, and 14, as well as the ePD, CNIL, IAPP, EDPB, and ICO were cited. 



\subsubsection{DPA} The posts within this category looked for insights regarding the agreement between data processors and data controllers, what to include in a DPA document, or when they are required. One post asked \emph{``if a DPA was required with their internet service provider for a self-hosted website, due to IP addresses counting as online identifiers.''} Another post by a B2B cloud software provider asked \emph{``if a DPA is necessary if their privacy policy and terms of service documents contain the same information.''} Replies to this post sought clarification about the nature of the business, but all mentioned that a separate DPA document was not necessary. Some replies highlighted that while the business needs an agreement with their customers (the data controllers), no agreement was necessary with users of the customer's websites. One reply pointed out the source of confusion as being a distinction between the GDPR's definition of a DPA (a specific document) and the colloquial use of the term to refer to any agreement to meet the requirements of GDPR Art. 28.

\subsubsection{Data Transfer} Posts in this category discussed the transfer of data to third countries, especially in the absence of any agreements. One post asked about \emph{``access to sensitive data with remote workers in other countries.''} The top response suggested that the poster conduct a risk assessment on the data and who currently has access. It also highlighted that additional steps would be needed if the remote workers are in non-EU countries. The original poster expressed confusion about whether there is a legal distinction between data at rest and data in transfer.
Several other posts asked about how they could use US services as an EU-based company. For example, one post listed US-based transactional services and asked if they can be used in a compliant manner. Another asked if their EU-based SaaS, which has exclusively North American customers, needed to comply with GDPR. A response pointed to Art. 3, as well as the EDPB guidelines on the territorial scope of the GDPR. A significant portion of the discussion on these posts involved the Schrems 2 case, which invalidated the EU-US Privacy Shield. Responses often conflicted, with responders pointing to information published both before and after this case, causing significant confusion.


\subsubsection{DPIA$^{*}$} Users also requested advice about DPIAs, including what these documents entail and how to assess risks prior to starting a project. The content of posts in this category varied significantly, but the general trend indicated that users were often unaware of when DPIAs were necessary. For example, a user planned to \emph{``implement real-time tracking mechanisms''} and sought advice on \emph{``how to clearly mitigate the risks prior to deployment.''} One reply simply advised the poster to conduct a formal risk assessment before continuing. Another response agreed with this suggestion, but also discussed the implications of such tracking, highlighting GDPR Art. 5, 11, 13, and 17. The same response pointed to both the EDPB/WP29 and the ICO guidelines, \emph{``even outside of the UK.''}
Another post discussed starting an E-learning forum and sought advice on logging practices prior to developing the software. The poster was uncertain whether they must log viewership details of each user, so that users may know who accessed their data. Responses to this post asked for further details about the goals of such logging, highlighting that there was no GDPR requirement for this use case. One reply specifically claimed that the opposite may be true, that \emph{``such data collection could easily be a GDPR violation.''}

Another post, by a user who stated they work at a privacy consulting firm, asked if a DPIA was necessary for the sensor data of a specific medical instrument that monitors patient movements, mentioning that their client is an ``elderly care house''. One response discussed their own related experience and broadly recommended that a DPIA be completed. Another asked for follow-up details, such as the country the poster was in, and referred to the Italian Supervisory Authority, which also recommends conducting a DPIA.

\subsubsection{DSAR$^{*}$} Concerns about DSARs revolved around situations where users request access to their data. One poster expressed confusion as to whether it includes more than just PII, asking \emph{``When a user requests a copy of their information, is this all data, or only PII?''} The responses were conflicting, with one user pointing to Google as a baseline, suggesting that \emph{``I think they have to give you everything.''} Another answered that only the personal data the user directly provides must be given, but nothing derivative from it. Several posts asked broadly how other community members were managing DSARs or which tools they were using to make handling requests easier. One common response when posters were asking about data deletion was to highlight that \emph{``the data deletion request right is not absolute''}, and that data retention is appropriate when there is a legal obligation to do so. One post by an employee of an educational platform for institutions asked about a deletion request they received directly from one of their client's users. The consensus among all responses was that it is the data controller's responsibility to handle such requests, not the data processor. Some replies linked to ICO guides for supporting information on individual rights and the obligations between controllers and processors.

\subsubsection{RoPA$^{*}$} A small portion of posts sought advice directly on RoPA documents, as well as what is considered PII requiring RoPA, platforms to manage RoPA, and on data retention requirements tied to RoPA.
One user asked \emph{``how to keep RoPA organized in a large organization''} and sought recommendations for \emph {``more efficient management solutions.''} Another asked \emph{``whether a referrer URL, user agent, and three-quarters of an IP address together count as personal data and whether the anonymization process falls under RoPA?''} In addition to Art. 30, many of the responses referred to GDPR articles such as Art. 13, 14, 17, and 28. For example, one response discusses \emph{``"You will have mapped all of the personal data you process to the purpose and lawful basis for your RoPA (art 30) and that will align with your privacy notice (art 13-14). It should be fairly easy to add your retention to the above.''} The discussions on RoPA and the number of articles to consider indicate its complexity. 

\subsubsection{Data Breach Notification$^{*}$} A few posts discussed the implications of data breaches, how they should be handled, what constitutes a data breach under GDPR, what the responsibilities of the third-party are, and when to report the breach. For example, one user asked about \emph{``Third-Party Servicer Compromised in Data Leak that (potentially) contained PII information on our clients, some of whom reside within UK/EU Citizen Information. Who discloses this to the GDPR Regulatory Authority? My company or the third-party servicer?''} The discussions revolved around the fact that it is the data controller's responsibility. Most posts had a follow-up question from the poster. In the case above, the poster also asked whether the number of affected people matters in this context. The response to what constitutes a data breach had some confusion, e.g., one responder mentioned \emph{``as the definition of personal data breach in the law appears to exclude temporary loss of availability of data. However, guidance from the ICO and EDPB both appear to include it.''} 

\subsubsection{GDPR Definitions \& Generic Compliance} Some users needed clarification on basic GDPR terminologies or expressed confusion about ensuring GDPR compliance. Posts in these two categories did not closely relate to other major challenges.
For example, many were unsure how to treat certain types of data under GDPR. One user asked about \emph{``which pieces of medical information needed to be handled differently within their app.''} Another sought clarification on the \emph{``differences between a data processor and a data controller.''} Regarding compliance, another post asked \emph{``if GDPR applied to their business if they use a third-party payment processor and do not collect any data themselves.''} Among these posts, responses often cited specific GDPR articles and EDPB guidelines for clarification.



\subsubsection{Others} 9\% of posts raise concerns not aligned with any of the challenge categories. These posts reflect diverse issues, like users promoting their own privacy tools or developers looking for resources to learn about privacy regulations. One post from an Australian developer \emph{``expressed frustration about the perceived jurisdictional overreach of GDPR, questioning the EU’s authority to regulate non-EU businesses and proposing to block EU users.''} 


\section{Discussion}
\label{sec:discussion}
\noindent\textbf{Key Insights and Implications.} Our findings suggest that practitioners use these forums to navigate privacy compliance challenges in practice. 
Although our survey was posted in EU-focused subreddits, most participants were from North America. Despite 62\% of survey respondents reporting they had in-house privacy counsel, $\sim$38\% either had counsel without privacy experience, no counsel, or were unsure. This dichotomy suggests several possibilities supported by a recent study of 51 US privacy attorneys~
\cite{NB26}: in-house self-service resources may be limited, do not exist, or are unsituated in design contexts; those with legal advice see opportunities to share their experiences with those who lack access to advice; and/or there is a translation challenge wherein access to legal advice does not guarantee that it can be easily or effectively translated into compliance design decisions. Our results indicate that while the presence of in-house legal teams increases confidence in companies' privacy practices \cite{prybylo2024evaluating, horstmann2024those, tahaei2021privacy}, participants continue to engage with external forums even when internal expertise is available.

Our analyses show there is high motivation to provide advice (83\%) and to share developer perspectives and experiences (82\%). This high motivation creates challenges in determining what information is most credible. Respondents report a variety of means to check the credibility of responses, including checking poster profiles (22\%), searching the web (18\%), studying the post thread for citations (17\%), among others. Overall, however, the means to assess credibility are ad hoc and rely heavily on features provided by Reddit (e.g., upvoting, threaded comments that may include critiques). $\sim$11\% report rarely checking the credibility. Despite the earlier result that most participants have in-house privacy counsel (62\%), or legal counsel in general (87\%), a key finding is the risk that developers may still encounter inaccurate or inconsistent information that is difficult to evaluate.

Finally, we found that discussions on regulatory-focused subreddits do cover permission-related issues, similar to prior work~\cite{parsons2023understanding,li2021developers, tahaei2020understanding}, and that consent, especially cookie consent, was the most prominent compliance challenge discussed. However, unlike prior work, we uncovered a broader set of challenges regarding consent, focusing mainly on design decisions, e.g., the need for a decline button, which design features to include, and when consent is required, which are motivated by nuances in shifting interpretations of privacy law. For example, the European Data Protection Board clarified in 2020 that declining cookies must be as easy as accepting cookies\footnote{Guidelines 05/2020 on consent under Regulation 2016/679, Adopted on 4 May 2020}, and in 2023 clarified that the reject cookies button must be prominent displayed\footnote{The report undertaken by the Cookie Banner Taskforce, Adopted on 17 Jan. 2023}, over five years after the GDPR became effective. Despite clarifications from regulators, these challenges persist, highlighting the need for more nuanced practical guidance.  

\textbf{Future Research Directions.}
Developers’ reliance on external advice may indicate a lack of in-house self-service resources for privacy-by-design (PbD) practices. If so, this creates an opportunity for the development of techniques that align with conventional legal advice, avoid dark patterns, and are grounded in design contexts. While regulators provide broad compliance guidance intended to address design questions, such guidance is often limited and insufficiently contextualized. Advances in LLMs have led to human-AI design ideation~\cite{SSW+25}, including the exploration of design alternatives beyond code generation~\cite{ZJT+25}. Investigating human-AI collaboration on PbD may support the need identified in our analysis. In addition, techniques to synthesize privacy requirements from laws could make regulatory information more accessible to developers, e.g., by extracting reusable requirements from privacy laws~\cite{baldwin2025prompts}.

Another key finding is the difficulty that users face when evaluating privacy advice in subreddits. Prior research has examined how general Internet users judge the credibility of web pages~\cite{FSD+03} and search results~\cite{YT11}, and evaluate health-related information~\cite{CZG21}. However, these sources are often manipulated to monetize the public, whereas other issues underpin the credibility of privacy advice, such as the advisers' access to reliable information and its comprehension, the shift in legal interpretation and regulator guidance, and clarity of communication between the poster and reader. New research that reconciles posted claims with broader legal contexts or estimates poster credibility based on message history (e.g., how often they are consistent with established facts) could provide new tools for fact-checking PbD advice from forums.


Focusing directly on communication, it is essential to examine closely how posters and responders surface assumptions and facts in the posts and resolve misunderstandings and ambiguities during discussions \cite{santos2024patterns}. 
Modeling these interaction patterns could support the design of dialogue policies (i.e., a set of rules to guide the models) for LLM-based privacy assistants and compliance tools~\cite{eberhart2021dialogue, eberhart2022generating}. 



\textbf{Educational Takeaways.}
Our findings show that users seek GDPR advice on forums, highlighting a gap in privacy compliance education and training, which is consistent with prior work~\cite{prybylo2024evaluating}. We observe that users are unfamiliar with GDPR requirements and guidelines, such as the EDPB’s Cookie Banner Report (1/18/23), which is cited infrequently. These results emphasize the need for \emph{companies} and \emph{educators} to create modules, courses, and training materials on topics such as consent, PII identification, data processing agreements, data transfers, and broader regulatory compliance. Our survey results show that having privacy-related experience positively impacts how users interact on forums and helps them better critically assess the credibility of responses; yet, a quarter of participants do not have internal corporate training. This gap illustrates how privacy professional trade groups, such as the International Association of Privacy Professionals, could contribute to ongoing professional learning for engineers on PbD topics.


\textbf{Implications for Industry and Regulators.} Accurately implementing cookie consent banners and obtaining consent are the top challenges identified in our findings. Thus, companies and open source developers could create Standard Development Kits (SDKs) with privacy-related UI components, such as cookie banners, alongside other reusable components that avoid dark patterns \cite{ahuja2026dark}. Automated tools that identify privacy-related source code and use machine learning to generate privacy notices could assist developers~\cite{jain2023, jain2026automated} if they were made more easily accessible through IDE plugins. The need for tools and templates for creating effective DPA, RoPA, and DPIA was a recurring question. Regulators could provide templates and practical examples for such documents, while companies could develop tools to (partially) automate the process.


%
\section{Conclusion}
\label{sec:conclusion}

In this paper, we conducted a quantitative survey of 223 users on regulatory-focused subreddits and a qualitative analysis of 2,248 posts using an LLM-based approach to understand users' professional expertise, common privacy challenges, and motivations and practices for seeking advice or responding on Reddit. We provide detailed analysis of GDPR-related challenges, the interaction and engagement behavior of Reddit users, credibility assessment strategies, and how they impact decision-making processes. We discuss future research and education directions based on our findings. 



\begin{acks}
The authors used generative AI-based tools to correct typos, grammatical errors, and awkward phrasing.

This research was funded by NSF CAREER \#2238047 and NSF Award \#2217572.
\end{acks}

\bibliographystyle{ACM-Reference-Format}
\bibliography{references}

\appendix

\section{Data Availability}
\label{sec:openscience}

The repository containing the list of survey questions and the Reddit post analysis, including the ground truth and labeled datasets, all prompts, scripts, sample data for prompts, and results, can be found here: https://github.com/PERC-Lab/GDPR\_discussion\_data. 





\section{Research Hypotheses and Results}
\label{app:Hypolist}



\subsection{Variability in Motivations for Participating in GDPR-related Discussions}
\label{app:H1list}

\begin{itemize}
    \item \textbf{H1a:} Users’ motivations for participating in GDPR-related discussions vary by job roles.
    \item \textbf{H1b:} Users’ motivations for participating in GDPR-related discussions vary by privacy-related certifications.
    \item \textbf{H1c:} Users’ motivations for participating in GDPR-related discussions vary by privacy-related experience.
    \item \textbf{H1d:} Users’ motivations for participating in GDPR-related discussions vary by company size.
\end{itemize}

Table \ref{table:H1_tests} summarizes the statistically significant results of the hypothesis analysis.

\begin{table*}[h]
\centering
\caption{Statistically significant results for Hypothesis H1a to H1d}
\begin{adjustbox}{max width=\textwidth}
\begin{tabular}{|l|l|c|c|c|c|}
\hline
\textbf{Factor} & \textbf{Motivation} & \textbf{Statistic} & \textbf{p-value} & \textbf{Effect Size} & Sample Size \\
\hline

Privacy Certifications &
Create posts to share knowledge/experience &
$U = 956.50$ &
0.001 &
$r = 0.476$ &
N = 94 \\
\hline

Privacy Certifications &
Create posts to express opinions &
$U = 859.50$ &
0.032 &
$r = 0.310$ &
N = 95\\
\hline

Privacy-related Experience &
Create posts to share knowledge/experience &
$H = 14.30$ &
0.006 &
$\epsilon^2 = 0.112$ &
N = 95\\
\hline

Privacy-related Experience &
Reply to share perspectives/experience &
$H = 8.80$ &
0.032 &
$\epsilon^2 = 0.051$ &
N = 117\\
\hline

Company Size &
Create posts to express opinions &
$H = 12.21$ &
0.016 &
$\epsilon^2 = 0.328$ &
N = 30\\
\hline

Company Size &
Create posts to seek advice/feedback &
$H = 10.96$ &
0.027 &
$\epsilon^2 = 0.279$&
N = 30 \\
\hline

Company Size &
Reply to share perspectives/experience &
$H = 9.02$ &
0.029 &
$\epsilon^2 = 0.140$&
N = 47 \\
\hline

\end{tabular}
\end{adjustbox}
\label{table:H1_tests}
\end{table*}

\subsection{Variability in GDPR Challenges}
\label{app:Hi2list}
\begin{itemize}
    \item \textbf{H2a:} The selection of specific GDPR challenges varies based on company type.
    \item \textbf{H2b:} The selection of specific GDPR challenges varies based on company size.
    \item \textbf{H2c:} The selection of specific GDPR challenges varies based on company location.
    \item \textbf{H2d:} The selection of specific GDPR challenges varies based on participants’ job roles.
    \item \textbf{H2e:} The selection of specific GDPR challenges varies based on participants’ privacy-related certifications.
\end{itemize}
Table \ref{tab:h2results} shows the results of the hypotheses analysis.

\begin{table}[H]
\centering
\caption{Results for Hypothesis H2a to H2e}
\begin{adjustbox}{max width=\columnwidth}
\begin{tabular}{|c|c|c|c|c|c|}
\hline
\textbf{} & \multicolumn{3}{|c|}{\textbf{Company}} & \textbf{Job Role} & \textbf{Pri-Certified} \\
\cline{2-4}
\textbf{} & \textbf{Type} & \textbf{Size} & \textbf{Location} & \textbf{} & \textbf{} \\
\hline

Statistic &
$\chi^2(24)$ = 59.3299 &
$\chi^2(12)$ = 24.1388 &
$\chi^2(36)$ = 44.0361 &
$\chi^2(28)$ = 27.2474 &
$\chi^2(4)$ = 26.1499 \\
\hline

p-value &
0.0001 &
0.0195 &
0.1681 &
0.5048 &
$< 0.001$ \\
\hline

Cramér’s V &
0.2123 &
0.1564 &
0.1832 &
0.1439 &
0.2819 \\
\hline

Sample Size &
190 &
190 &
189 &
190 &
190 \\
\hline

\end{tabular}
\end{adjustbox}
\label{tab:h2results}
\end{table}

\subsection{Influence of Legal Support on Compliance Confidence}
\label{app:Hi3list}
\begin{itemize}
  \item \textbf{H3a:} Organizations with in-house privacy legal teams report higher confidence in privacy compliance.
    \item \textbf{H3b:} Organizations that frequently seek external legal advice report lower confidence in privacy compliance.
\end{itemize}
Table \ref{table:H2l_pvalues} shows the results of the hypotheses analysis.

\begin{table}[H]
\centering
\caption{Results for Hypothesis H3a and H3b}
\begin{adjustbox}{max width=\columnwidth}
\begin{tabular}{|l|c|c|}
\hline
 & \textbf{In-house Legal Team} & \textbf{External Legal Advice} \\
\hline
Statistic & $H$ = 8.4773 & $\rho$ = 0.0083 \\
\hline
P-values & 0.0371 & 0.9096 \\
\hline
Effect Size & $\eta^2$ = 0.0294 & Negligible \\
\hline
Sample Size & 190 & 191 \\
\hline
\end{tabular}
\end{adjustbox}
\label{table:H2l_pvalues}
\end{table}

\subsection{GDPR Knowledge Sources}
\label{app:Hi4list}
\begin{itemize}
    \item \textbf{H4a:} Participants’ primary sources of GDPR knowledge vary by job roles.
    \item \textbf{H4b:} Participants’ primary sources of GDPR knowledge vary by company size.
    \item \textbf{H4c:} Participants’ primary sources of GDPR knowledge vary by educational background.
    \item \textbf{H4d:} Participants’ primary sources of GDPR knowledge vary based on privacy-related certifications.
\end{itemize}
Table \ref{table:H3_pvalues} shows the results of the hypotheses analysis.

\begin{table}[h]
\centering
\caption{Result for Hypothesis H4a and H4d}
\begin{adjustbox}{max width=\columnwidth}
\begin{tabular}{|l|c|c|c|c|}
\hline
\textbf{} & \textbf{Job Role} & \textbf{Company Size} & \textbf{Education} & \textbf{Pri-Certified} \\
\hline

Statistic &
$\chi^2(56)$ = 31.24 &
$\chi^2(32)$ = 18.58 &
$\chi^2(32)$ = 15.09 &
$\chi^2(8)$ = 22.19 \\
\hline

p-value &
0.9970 &
0.9717 &
0.9951 &
0.0046 \\
\hline

Cramér’s V &
0.0830 &
0.1290 &
0.1163 &
0.2820 \\
\hline

Sample Size &
191 &
68 &
68 &
191 \\
\hline

\end{tabular}
\end{adjustbox}
\label{table:H3_pvalues}
\end{table}



\subsection{Credibility Evaluation by Professional Background}
\label{app:H6list}
\begin{itemize}
  \item \textbf{H5a:} Participants’ credibility evaluation strategies are associated with their job role.
  \item \textbf{H5b:} Participants’ credibility evaluation strategies are associated with their educational background.
  \item \textbf{H5c:} Participants’ credibility evaluation strategies are associated with their company size.
  \item \textbf{H5d:} Participants’ credibility evaluation strategies are associated with their privacy-related certifications.
\end{itemize} 
Table \ref{table:Q25_anova_pvalues} shows the results of the hypotheses analysis.

\begin{table}[h]
\centering
\caption{Results for Hypothesis H5a to H5d}
\begin{adjustbox}{max width=\columnwidth}
\begin{tabular}{|p{2.4cm}|c|c|c|c|}
\hline
\textbf{} & \textbf{Job Role} & \textbf{Company Size} & \textbf{Education} & \textbf{Pri-Certified} \\
\hline

Main Effect (Group) &
$F(7,86)=2.12$ &
$F(4,24)=2.00$ &
$F(4,24)=0.52$ &
$F(1,92)=0.57$ \\
\hline

$p$-value (Group) &
0.0498 &
0.1260 &
0.7186 &
0.4521 \\
\hline

$\eta_p^2$ (Group) &
0.147 &
0.250 &
0.080 &
0.006 \\
\hline

Interaction &
$F(42,516)=1.27$ &
$F(24,144)=2.43$ &
$F(24,144)=0.60$ &
$F(6,552)=0.80$ \\
\hline

$p$-value (Interaction) &
0.1270 &
$<$0.001 &
0.9251 &
0.5692 \\
\hline

$\eta_p^2$ (Interaction) &
0.093 &
0.288 &
0.091 &
0.009 \\
\hline

Sample Size &
94 &
29 &
29 &
94 \\
\hline

\end{tabular}
\end{adjustbox}
\label{table:Q25_anova_pvalues}
\end{table}


\section{Participants' Demographic Information}
\label{app:demo}
Table \ref{table:demographics2} shows the demographic information of the participants.

\begin{table*}[h]
    \centering
    \footnotesize
    \caption{Demographic Information about the Participants}
    \renewcommand{\arraystretch}{1.2}
    \resizebox{\textwidth}{!}{
        \begin{tabular}{|l|l|}
            \hline
            \textbf{Category} & \textbf{Details} \\ 
            \hline
            \textbf{Gender} & Female (26\%), Male (73\%), Non-binary (1\%) \\ 
            \hline
            \textbf{Age} & 18-24 (8\%), 25-34 (40\%), 35-44 (46\%), 45-64 (6\%) \\ 
            \hline
            \textbf{Education} & High School (3\%), Associate (4\%), Undergraduate (21\%), Graduate (34\%), Professional (39\%) \\ 
            \hline
            \textbf{Degree} & CS (35\%), Data Science (22\%), IT (21\%), Business (19\%), Other (6\%) \\ 
            \hline
            \textbf{Company Size} & 1-5 (4\%), 6-20 (16\%), 21-50 (28\%), 51-100 (23\%), 101+ (30\%) \\ 
            \hline
        \end{tabular}
        }
\label{table:demographics2}
\end{table*}

\section{Distribution of Participants' Roles}
\label{app:jobrole}
\begin{figure}[h]
    \centering
    \begin{subfigure}{0.48\textwidth}
        \centering
        \includegraphics[width=\textwidth]{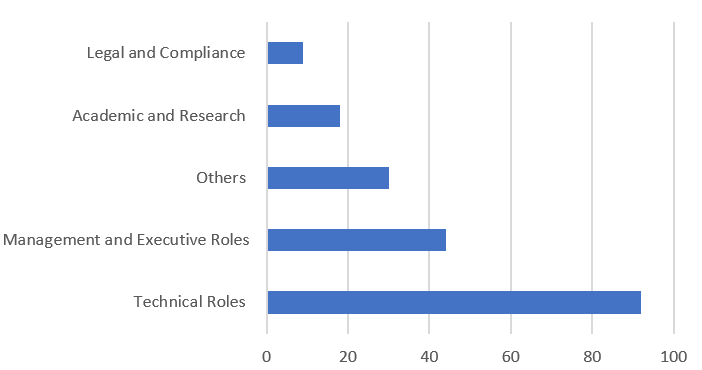}
        \caption{Distribution of the Job Roles}
        \label{fig:Annotated job title}
    \end{subfigure}
    \hfill
    \begin{subfigure}{0.48\textwidth}
        \centering
        \includegraphics[width=\textwidth]{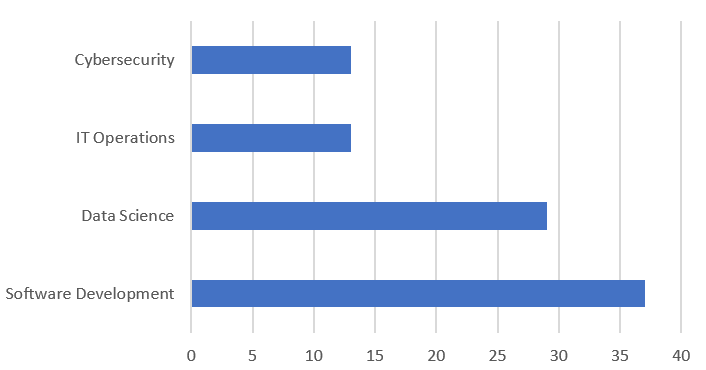} 
        \caption{Top Four Technical Roles Subcategories }
        \label{fig:Technical Roles subcategories}
    \end{subfigure}
    \caption{Distribution of Participants' Roles}
\end{figure}

Figure~\ref{fig:Annotated job title} and Figure~\ref{fig:Technical Roles subcategories} show the distribution of job roles and the subcategories within technical roles.

\section{Distribution of Privacy Certification Types}
\label{app:cert}
\begin{figure}[h]
     \centering
        \includegraphics[width=0.42\textwidth]{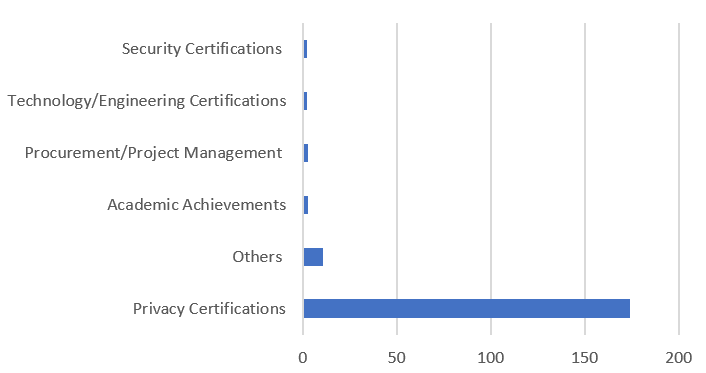} 
        \caption{Categories of Privacy Certification Types}
        \label{fig:Q6 Responses}
\end{figure}

Figure~\ref{fig:Q6 Responses} shows privacy-related certifications dominate the responses.

\section{Frequency of GDPR-related Engagements}
\label{app:Frequencyfig58}
\begin{figure}[h]
    \centering
    \includegraphics[width=0.4\textwidth]{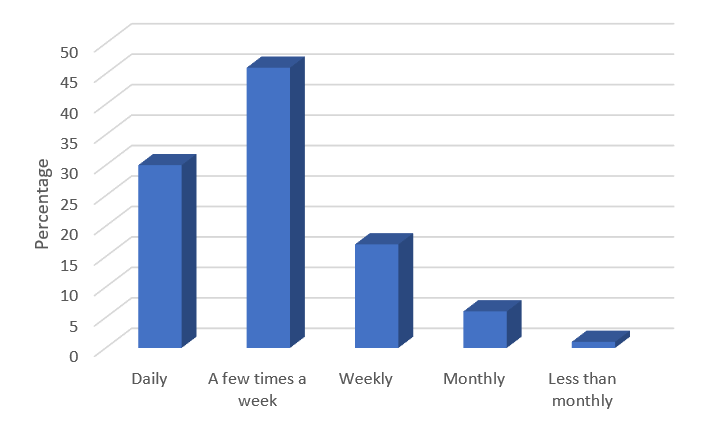} 
    \caption{Frequency of GDPR-related Engagements}
    \label{fig:58}
\end{figure}

Figure~\ref{fig:58} shows the GDPR-related engagements.

\section{Frequency of Posting or Replying}
\label{app:Frequencysixmonths}
\begin{figure}[h]
    \centering
    \includegraphics[width=0.4\textwidth]{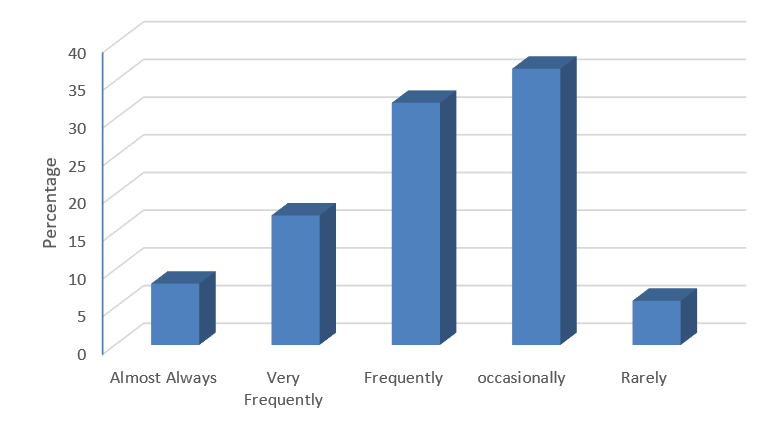} 
    \caption{Frequency of GDPR-related Posting/Replying}
    \label{fig:Fr6months}
\end{figure}

Figure~\ref{fig:Fr6months} shows the frequency of users’ posting and replying activity on GDPR-related Reddit posts during the last six months.

\section{Frequencies of Interaction Types on Reddit}
\label{app:Frequencys}
Table~\ref{tab:reddit_interaction} shows the frequency of users’ interaction types on Reddit.

\begin{table}[h]
\centering
\caption{Frequencies of Interaction Types on Reddit}
\begin{tabular}{|p{5.8cm}|c|}
\hline
\textbf{Interaction Type (Q17)}  & \textbf{\% (Count)}  \\ \hline
Primarily create original posts  & 19\% (41) \\ \hline
Mostly reply to other users' posts   & 31\% (68) \\ \hline
Equally create original posts and reply  & 34\% (75)\\ \hline
Mostly read posts w/o posting or replying  & 17\% (37)  \\ \hline
\end{tabular}
\label{tab:reddit_interaction}
\end{table}

\section{Use of External Privacy or Legal Advice}
\label{app:q12}
\begin{figure}[bt]
    \centering
    \includegraphics[width=0.42\textwidth]{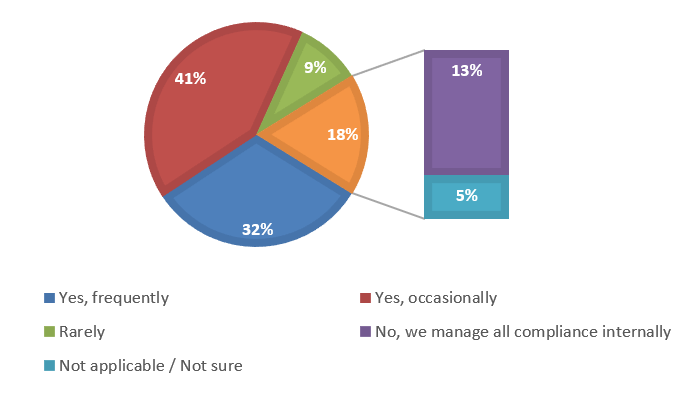} 
    \caption{Use of External Privacy or Legal Advice}
    \label{fig:Q12}
\end{figure}

Figure~\ref{fig:Q12} shows the distribution of participants’ use of external privacy or legal advice.

\section{Shared GDPR Challenges in Survey vs. Posts}
\label{app:shared_gdpr_challenges}
\begin{figure}[ht]
    \centering
    \includegraphics[width=0.4\textwidth]{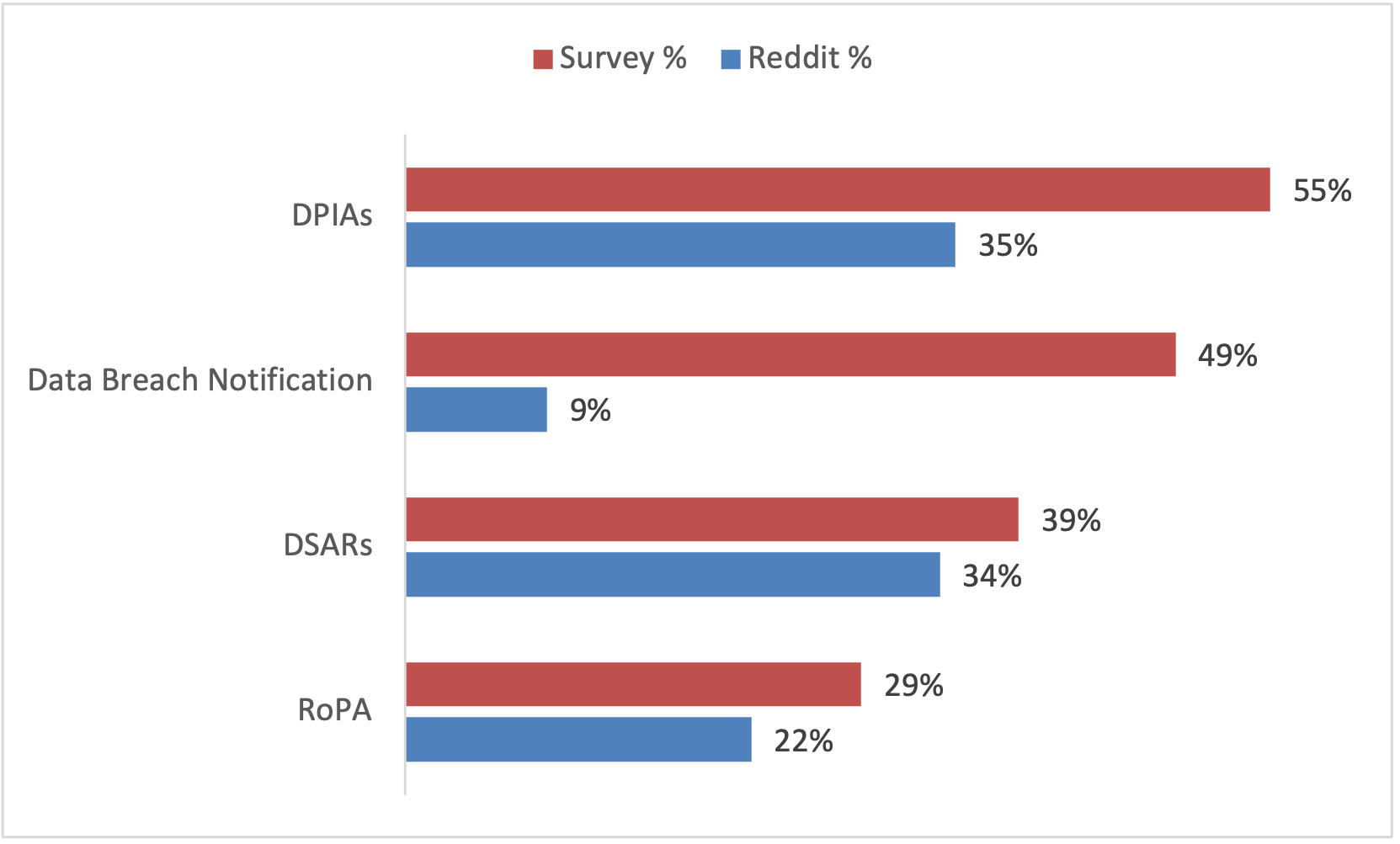}
    \caption{Shared GDPR Challenges in Survey vs. Posts}
    \label{fig:reddit_analysis_label_exist_dist}
\end{figure}

Figure~\ref{fig:reddit_analysis_label_exist_dist} shows the proportion of GDPR challenges present in both the survey and Reddit post analysis.

\section{Reddit Posts Analysis}
\label{app:redditpostanalysis}

\subsection{Privacy Terms for Reddit Queries}
\label{app:privacykeywords}
Table~\ref{tab:privacy_keywords} shows the terms extracted from the Hark methodology and their expanded synonyms. Some terms were removed from the final set, but since their synonyms were retained, we include them.

\begin{table}[H]
\centering
\caption{Privacy-related Terms used for Reddit Queries}
\label{tab:privacy_keywords}
\renewcommand{\arraystretch}{1.2}
\begin{tabular}{|p{2.7cm}|p{4.8cm}|}
\hline
\textbf{Harks Terms \cite{harkous2022hark}} & \textbf{Added Synonym Terms} \\
\hline
Privacy & --- \\
\hline
Surveillance & Monitoring, Tracking \\
\hline
Aggregation & Collection, Use, Processing, Sharing, Transfer \\
\hline
Identification & Authentication, Profiling \\
\hline
\st{Secondary Use}  & Data Repurposing \\ 
\hline
Exclusion & --- \\
\hline
Breach & Vulnerability, Compromise, Unauthorized, Access, Exposure, Leak \\
\hline
Confidentiality & Anonymity \\
\hline
Disclosure & --- \\
\hline
Notice & --- \\
\hline
Data Minimization & Data Protection \\
\hline
Purpose Specification & --- \\
\hline
\st{Use Limitation}  & Access Control \\
\hline
Choice & --- \\
\hline
Consent & Permission, Authorization, Agreement, Opt-in, Opt-out, Retention \\
\hline
\end{tabular}
\end{table}

\subsection{Ground Truth Creation}
\label{app:labelingsoftwareposts}

Table~\ref{tab:annotation_summary_reddit_post} shows the results for each round of annotation when identifying posts related to software development and privacy.

\begin{table}[h]
    \centering
    \caption{Annotation Summary for Relevant Posts}
    \footnotesize
    \renewcommand{\arraystretch}{1.2}
    \resizebox{\columnwidth}{!}{%
    \begin{tabular}{|>{\centering\arraybackslash}p{1.0cm}|
                    >{\centering\arraybackslash}p{1.4cm}|
                    >{\centering\arraybackslash}p{1.6cm}|
                    >{\centering\arraybackslash}p{1.4cm}|
                    >{\centering\arraybackslash}p{1.4cm}|}
        \hline
        \textbf{Rounds} & \textbf{Posts Reviewed} & \textbf{Disagreements} & \textbf{Cohen’s Kappa} & \textbf{Relevant Posts} \\
        \hline
        Round 1 & 100 & 11 & 0.749 & 32 \\
        \hline
        Round 2 & 100 & 14 & 0.684 & 32\\
        \hline
        Round 3 & 100 & 9 & 0.808 & 34\\
        \hline
        Round 4 & 50 & 4 & 0.834& 18\\
        \hline
        \textbf{Total} & \textbf{350} & & & \textbf{116} \\
        \hline
    \end{tabular}
    }
    \label{tab:annotation_summary_reddit_post}
\end{table}

\subsection{Prompt Strategies and Results}
\label{app:promptsengineering}



\paragraph{Privacy and Development Labels.} We developed parts of our CoT prompts following the guidelines described in \cite{tornberg2024textannotation, zadenoori2025prompts}. 
We accessed open-source models via Hugging Face~\cite{huggingfaceHuggingFace} and OpenAI models through its API~\cite{openaiPlatform}. For all models except GPT-5 and GPT-5-mini, we set $temperature = 0$ and $top\text{-}p = 1.0$. For GPT-5 and GPT-5-mini, we set the verbosity and reasoning effort to ``medium,'' and $top-p = 1.0$.

Listing \ref{listing:cot-dev-privacy} shows the CoT prompt used for detecting privacy and development posts. Note that the prompts for zero-shot and few-shot are shown on our GitHub repository.

\begin{lstlisting}[language=HTML,  caption={CoT Prompt for Privacy \& Development Posts},label={listing:cot-dev-privacy}, 
basicstyle=\scriptsize\ttfamily,
breaklines=true,
]
You are an expert system that needs to determine if Reddit posts are related to {topic} with either True or False and to provide an explanation for your reasoning.
Let's analyze how to classify posts step by step:
Step 1: Review the examples of posts and their answers.
Example 1: {question}
Post: {example 1}
Explanation: {reasoning for example 1}
Answer: {label of example 1}
Example 2: {question}
Post: {example 2}
Explanation: {reasoning for example 2}
Answer: {label of example 2}

Step 2: Understand how to apply this reasoning to new posts, using the examples and explanations as guidance.

Step 3: Apply this understanding to the following question and post.

Step 4: Determine the answer to this question for the following post. Make sure to review your response and provide a rationale for your selection:
{question}
Post: {post text}
Explanation:
\end{lstlisting}



We instantiated \{question\} from the two queries in Listing \ref{listing:question-privacy-dev}.

\begin{lstlisting}[language=HTML, caption={Questions Used in the Prompt for Privacy \& Development Detection},label={listing:question-privacy-dev}, basicstyle=\scriptsize\ttfamily,
breaklines=true,
]
Privacy Question: Is this Reddit post related to privacy or data protection issues under any context, including personal data, GDPR, or online privacy? Posts that are surveys are completely unrelated. Answer only with True or False.

Development Question: Is this Reddit post related to software, software development, developers, programming, engineering practices, software development activities, or code-related issues? Software development includes requirements analysis, software specification, design and development, quality assurance, testing, and maintenance. Posts that are surveys are completely unrelated. Answer only with True or False.
\end{lstlisting}


Table~\ref{tab:merged_performance} compares the performance of various models in terms of accuracy, precision, recall, and F1 score on detecting privacy and development related posts using CoT prompting on test dataset.

\begin{table}[H]
\centering
\caption{Performance on Detecting Privacy and Development Related Posts with Chain-of-Thought Prompting on Test Set}
\label{tab:merged_performance}
\renewcommand{\arraystretch}{1.2}
\resizebox{\columnwidth}{!}{
\begin{tabular}{|p{2.5cm}|c|c|c|c|c|}
\hline
\textbf{Model} & \textbf{Topic} & \textbf{Accuracy} & \textbf{Precision} & \textbf{Recall} & \textbf{F1 Score} \\
\hline
Llama3.2-1B-Instruct & Privacy     & 0.7500 & \textbf{0.9811} & 0.7591 & 0.8560 \\
                     & Development & 0.4286 & 0.3034 & 0.6000 & 0.4030 \\
\hline
Llama3.2-3B-Instruct & Privacy     & 0.7071 & 0.9800 & 0.7153 & 0.8270 \\
                     & Development & 0.6071 & 0.4000 & 0.4444 & 0.4211 \\
\hline
Llama3.1-8B-Instruct & Privacy     & 0.8500 & 0.9754 & 0.8686 & 0.9189 \\
                     & Development & 0.7571 & 0.6897 & 0.4444 & 0.5405 \\
\hline
Qwen2.5-7B-Instruct  & Privacy     & 0.7714 & 0.9730 & 0.7883 & 0.8710 \\
                     & Development & 0.7643 & \textbf{0.9286} & 0.2889 & 0.4407 \\
\hline
GPT-3.5-turbo-1106   & Privacy     & 0.9429 & 0.9568 & 0.9429 & 0.9498 \\
                     & Development & 0.7429 & 0.7574 & 0.7429 & 0.6477 \\
\hline
gpt-4o-mini          & Privacy     & \textbf{0.9500} & 0.9570 & \textbf{0.9500} & \textbf{0.9535} \\
                     & Development & 0.4714 & 0.6592 & 0.4714 & 0.4552 \\
\hline
gpt-5-mini           & Privacy     & 0.9429 & 0.9568 & 0.9429 & 0.9498 \\
                     & Development & 0.8071 & 0.8022 & 0.8071 & 0.8008 \\
\hline
gpt-5                & Privacy     & 0.9286 & 0.9565 & 0.9286 & \textbf{0.9423} \\
                     & Development & \textbf{0.8571} & 0.8556 & \textbf{0.8571} & \textbf{0.8531} \\
\hline
\end{tabular}
}
\end{table}


\paragraph{GDPR Challenges Labels.} We randomly sampled 84 posts (i.e., development set) from the subset labeled with challenges to refine our prompt (i.e., zero-shot, one-shot, and CoT). The prompts were designed to yield only the most relevant challenge in each post to improve reliability. A generated label was counted as a true positive if it matched at least one label assigned to the post in the ground truth dataset; if the label did not match, then it was counted as a false positive; and if no label was generated, then it was counted as a false negative. Similar to the first task, we aim to maximize recall and minimize false negatives, since correcting labels is an easier task than identifying missing labels \cite{Jain2022}.





We instantiated {labels' list} from Table ~\ref{tab:GDPR_challenege_def} and provided their corresponding definitions. This list covers all categories except ``data transfer'' and ``generic compliance,'' which were introduced later after the LLM had classified a substantial portion of data under the category ``Others.''





Table~\ref{tab:challengestestset} shows the results of detecting GDPR challenges of posts with CoT Prompting on the test set.

\begin{table}[h]
\centering
\caption{Performance on Detecting GDPR-related Challenges using Chain-of-Thought Prompting on Test Set}
\label{tab:challengestestset}
\renewcommand{\arraystretch}{1.2}
\resizebox{\columnwidth}{!}{
\begin{tabular}{|p{2.0cm}|c|c|c|c|}
\hline
\textbf{Model} & \textbf{Accuracy} & \textbf{Precision} & \textbf{Recall} & \textbf{F1 Score} \\
\hline
Llama3.2-1B-Instruct & 0.3387 & 0.5527 & 0.3750 & 0.3584 \\
\hline
Llama3.2-3B-Instruct & 0.3387 & 0.4159 & 0.3438 & 0.2939 \\
\hline
Llama3.1-8B-Instruct & 0.4677 & 0.5353 & 0.4844 & 0.4768 \\
\hline
Qwen2.5-7B-Instruct  & 0.6452 & 0.6772 & 0.6719 & 0.6388 \\
\hline
GPT-3.5-turbo-1106   & 0.4677 & 0.4193 & 0.4688 & 0.4019 \\
\hline
gpt-4o-mini          & 0.6613 & 0.7513 & 0.6875 & 0.6683 \\
\hline
gpt-5-mini           & \textbf{0.7258} & \textbf{0.7811} & \textbf{0.7377} & \textbf{0.7357} \\
\hline
gpt-5                & 0.6774 & 0.7149 & 0.6875 & 0.6857 \\
\hline
\end{tabular}
}
\label{table:challengestestset}
\end{table}

Table \ref{tab:llm_labeling_results} shows the frequency of challenges labeled by GPT-5-mini, the number of false positive labels, and the final count after the manual review.

\begin{table}[h]
\centering
\caption{Frequency of Labeling Challenges with GPT-5-mini and Final Frequency after Review}
\label{tab:llm_labeling_results}
\renewcommand{\arraystretch}{1.2}
\resizebox{\columnwidth}{!}{
\begin{tabular}{|p{2.4cm}|c|c|c|}
\hline
\textbf{Challenge} & \textbf{LLM Predictions} & \textbf{False Positive} & \textbf{Final Count} \\
\hline
Others & 51 & 6 & 133 \\
\hline
Cookie-Consent & 200 & 19 & 108 \\
\cline{1-1}\cline{4-4}
General-Consent &  &  & 88 \\
\hline
DPA & 94 & 35 & 70 \\
\hline
Definition & 82 & 50 & 38 \\
\hline
DPIA & 22 & 10 & 22 \\
\hline
DSAR & 30 & 6 & 28 \\
\hline
RoPA & 21 & 7 & 17 \\
\hline
Breach-notification & 12 & 5 & 8 \\
\hline
\textbf{Total} & 512 & 138 & 512 \\
\hline
\end{tabular}
}
\end{table}

\section{GDPR Challenge Categories and Definitions}
\label{app:GDPRChallenegedef}

Table~\ref{tab:GDPR_challenege_def} defines the GDPR-related challenge categories used in the annotation process, including definitions, relevant legal basis, examples, and clarification notes of each category.

\begin{table*}[h]
    \centering
    \caption{Codebook for GDPR Challenge Categories with Definitions and Examples}
    \footnotesize
    \newcolumntype{L}[1]{>{\raggedright\arraybackslash}p{#1}}
    \renewcommand{\arraystretch}{1.2}
    \resizebox{\textwidth}{!}{%
    \begin{tabular}{|L{1.5cm}|p{7cm}|L{1cm}|p{3.5cm}|p{1cm}|}
        \hline
        \textbf{Category}& \textbf{Definition} & \textbf{Legal Basis} & \textbf{Examples} & \textbf{Source}\\
        \hline
        \textbf{Data Protection Impact Assessments (DPIAs)} & Covers posts about Data Protection Impact Assessments (DPIAs). This is a risk assessment done before starting a project by the data controller that may affect people's privacy. It helps identify and mitigate potential risks associated with the use and processing of large volumes of personal data. Note that, DPIA is a privacy impact assessment document. & Article 35 & ``What are the risks of X?'' \newline ``Is it safe to use X if you're an EU company?'' & Survey\\
        \hline     
        \textbf{Data Breach Notification} & Covers posts about data breach incidents. When personal data/information or PII is lost, stolen, or exposed, a breach occurs. Thus, the organization or data controller must notify data protection authorities within 72 hours of discovering a breach involving personal data.  & Article 33 &     ``Was this incident a breach?'' \newline
        ``As a third party, if I were aware of a breach, should I report it?'' & Survey\\        
        \hline
        \textbf{Data Subject Access Requests (DSARs)} & Covers posts about Data Subject Access Requests (DSARs). These allow individuals to request access to their personal data collected by the organization. The organization or a data controller must reply within a short time and include all relevant information. The DSAR also includes other rights as follows: the right to erasure (right to be forgotten), the right to access by the data subject, the right to rectification, and the right to object. &Articles 15–17 and 21–22 &
        ``We received a data subject request asking us to X.'' \newline
        ``What information should be included in a DSAR response?'' \newline
        ``Can I request to delete all my emails after leaving the company?'' & Survey\\
        \hline
        \textbf{Records of Processing Activities (RoPA)} & Covers posts about Records of Processing Activities (RoPA). Based on RoPA, organizations or data controllers must keep detailed records of what personal data they process, for what purpose, how it’s stored, and who it’s shared with. This document is completed after the data collection and processing is done. & Article 30 & ``Does collecting X mean we must keep processing records?'' & Survey\\
        \Xhline{1.0pt} 
        \textbf{Consent} & Covers posts that discuss whether it is required to obtain explicit or implicit consent or permission before collecting or processing personal data. Consent can be of type privacy policies,  privacy notices, cookie notices or cookie banners. Parental consent, cookie consent, etc, are also included. Note that consent is not about data subject access rights/requests or other GDPR compliance. & Articles 6–8 & ``Do I need consent to do X?'' \newline ``How do I ask for user consent before processing data?'' & During Ground Truth Creation\\
        \hline
        \textbf{Definitions} & Covers posts that ask for clarification or explanations of terminology. Definition refers to defining what it means by personal information, data processor, data controller, what is PII, etc. & Article 4 & ``Is an IP address considered personal data under GDPR?'' \newline ``Is a customer ID pseudonymised data under GDPR?'' & During Ground Truth Creation\\
        \hline
        \textbf{Data Processing Agreements (DPAs)} & Covers posts about Data Processing Agreements (DPA) which is a legally binding contract between the data controller (e.g., company) and the data processor (e.g., a third party) for processing data, duration of the processing, the nature and purpose of the processing, the type of personal data and categories of data subjects and the obligations and rights of the controller. For example, when the data processor is an email client, a cloud storage service, or website analytics software, you must have a data processing agreement with each of these services. Note that DPA is an agreement document. & Article 28(3) & ``Do I need a DPA with my cloud provider?'' \newline ``What should be included in a DPA?'' & During Ground Truth Creation\\
        \Xhline{1.0pt} 
        \textbf{Data Transfer} & Covers posts about transferring personal data to third countries (especially outside the EU) or to international organizations. & Articles 44-50 & ``How can I send PII to a US-based server?'' \newline ``Does Google Font send users' PII outside the EU?'' & After Full Dataset Labeled\\
        \hline
        \textbf{Generic Compliance} & Posts related to concerns surrounding simply complying with the GDPR, but lacking any specific relation to the other challenges. &  & ``Does it comply with GDPR?'' & After Full Dataset Labeled\\
        \hline
        \textbf{Others} & Posts that do not fit any of the above categories. & &``Why does the GDPR apply to my company if it is established outside the EU?''\newline ``Are EU regulators allowed to conduct spot checks without prior notice?''\newline ``Does GDPR affect VPN usage under Swedish law?''\newline ``We created a privacy preserving tool.'' & After Full Dataset Labeled\\
        \hline        
    \end{tabular}
    }
    \label{tab:GDPR_challenege_def}
\end{table*}

\end{document}